\documentclass{aa}

\usepackage{color}

\usepackage{graphicx}
\usepackage{txfonts}
\usepackage{natbib}

\bibpunct{(}{)}{;}{a}{}{,} 
\begin{document}


\title{Modified viscosity in accretion disks. Application to Galactic black hole binaries, intermediate
mass black holes, and AGN}
\author{ Miko\l{}aj Grz\k{e}dzielski\inst{1}, Agnieszka Janiuk\inst{1},
Bo\.zena Czerny\inst{1}
\and
Qingwen Wu \inst{2}
}

\institute{Center for Theoretical Physics,
Polish Academy of Sciences, Al. Lotnikow 32/46, 02-668 Warsaw, Poland \\
\email{mikolaj@cft.edu.pl}
\and
School of Physics, Huazhong University of 
Science and Technology, Wuhan 430074, China \\
}

   \date{Received ...; accepted ...}
   
  \abstract{\textit{Aims}
  {
 Black holes (BHs) surrounded by accretion disks are present in the Universe at different scales of masses, from microquasars up to the active
 galactic nuclei (AGNs). Since the work of Shakura and Sunyaev (1973) and their $\alpha$-disk model, various prescriptions for
 the heat-production rate are used to describe the accretion process. The current picture remains ad hoc due the 
 complexity of the magnetic field action. In addition, accretion
 disks at high Eddington rates can be radiation-pressure dominated and, according to some of the heating prescriptions, thermally unstable.
 The observational verification of their resulting variability 
patterns may shed light on both the role of radiation pressure and magnetic fields in the accretion process.}
 \\
  \textit{Methods}
   {
   We compute the structure and time evolution of an accretion disk, using the 
code GLADIS (which models the global accretion disk instability). We supplement this model with a modified viscosity prescription, which can to some extent 
 describe the magnetisation of the disk.
 We study the results for a large grid of models, to cover the whole parameter space, and we derive conclusions separately for different scales of black hole 
masses, which are characteristic for various types of cosmic sources. We show 
 the dependencies between the flare or outburst duration, its amplitude, and period, on the accretion rate and viscosity scaling. 
   }
\\
\textit{Results}
{
We present the results for the three grids of models, designed for different black hole systems (X-ray binaries, intermediate mass black holes, and galaxy centres).
 We show that if the heating rate in the accretion disk grows more rapidly with the total pressure and temperature, the instability results in longer and sharper flares. In general, we confirm that the disks around the supermassive black holes are more radiation-pressure dominated and present relatively
 brighter bursts.  Our method can also be used as an independent tool for the black hole mass determination, which we confront now for the intermediate black hole in the source HLX-1.
We reproduce the light curve of the HLX-1 source. We also compare the duration times 
 of the model flares with the ages and bolometric luminosities of AGNs. 
}
\\
\textit{Conclusions}
{
 With our modelling, we justify the modified $\mu$-prescription for 
 the stress tensor $\tau_{\rm r \phi}$ in the accretion flow in microquasars. 
The discovery of the Ultraluminous X-ray source HLX-1, claimed to be an intermediate black hole, gives further support to this result.
The exact value of the $\mu$ parameter, as fitted to the observed light curves, may be treated as a proxy for the magnetic field strength in the accretion flow in particular sources, or their states.}}
  \authorrunning{Grzedzielski et al.}
  \titlerunning{Modified viscosity}
   \maketitle
\section{Introduction}
\label{sect:intro}
Accretion disks are ubiquitous in the astrophysical black holes (BHs) environment,
and populate a large number of known sources.  Black hole masses range from
stellar mass black holes in X-ray binaries, through intermediate mass black holes (IMBHs), up to the supermassive blackholes
in quasars and active galaxy centers (Active Galactic Nuclei, AGNs). The geometrically thin, optically thick accretion disk that is described by 
the theory of Shakura and Sunyaev is probably most relevant for the high/soft spectral states of black hole X-ray binaries, as well as for some active 
 galaxies, such as Narrow Line Seyfert 1s and numerous radio quiet quasars \citep{brandt1997,peterson2000,foschini2015}. 
The basic theory of a geometrically thin stationary accretion 
is based on the simple albeit powerful $\alpha$ prescription for the viscosity in the accreting plasma,
introduced by \citet{1973A&A....24..337S}.
This simple scaling of viscous stress with pressure is also reproduced in the more recent numerical 
simulations of magnetised plasmas \citep{HiroseKrolikStone,2013ApJ...778...65J, Mishra2016}. However, the latter are still not capable of modelling the global dynamics, time variability, and radiation emitted in the cosmic sources, and hence
cannot be directly adopted to fit the observations.
The global models, however, must go beyond the stationary model, as 
the time-dependent effects connected with non-stationary accretion are clearly important.
In particular, a number of observational facts support the idea of a cyclic activity in the high-accretion-rate sources. One of the best studied examples is
the microquasar GRS 1915+105, which in some spectral states exhibits 
cyclic flares of its X-ray luminosity, well fitted to the
 limit cycle oscillations 
of an accretion disk on timescales of tens or hundreds of seconds \citet{GRS1997Taam,GRS2000Belloni,GRS2011Neilsen}.
Those heartbeat states are known since 1997, when the first XTE PCA observations of this 
source were published \citep{GRS1997Taam}, while recently 
yet another microquasar of that type, IGR J17091-3624, was discovered \citep{Revnivtsev,Kuulkers,Capitanio09}; 
heartbeat states were also found for this source \citep{Altamirano11a,Capitanio12,Pahari14,Janiuk2015}.
Furthermore, a sample of sources proposed in \citet{Janiuk2011} was 
suggested to undergo luminosity oscillations, possibly induced by the non-linear dynamics of the emitting gas. This suggestion was confirmed by the recurrence 
analysis of the observed time series, presented in \citet{Sukova2016}.
One possible driver of the non-linear process in the accretion disk
is its thermal and viscous oscillation induced by the radiation pressure term; it can be dominant for high enough accretion rates in the innermost regions of the accretion disk, which are the hottest.
The timescales of such oscillations depend on the black hole mass, and are on the order of tens to hundreds of seconds
for stellar mass BH systems.
For a typical supermassive  black hole of $10^{8} $M$_{\odot}$, 
the process would require timescales of hundreds of years.
 Therefore, in active galactic nuclei (AGNs) we cannot observe the evolution under the radiation pressure instability directly.
Nevertheless, statistical studies may shed some light on the sources' evolution.
For instance, the Giga-Hertz Peaked quasars \citep{Czerny2009} have very compact 
sizes, which would directly imply their ages. In the case of a limit-cycle kind of evolution, these sources would in fact not be very 
young, but `reactivated'.
Another observational hint is the shape of
distortions or discontinuities in 
the radio structures. These structures may reflect the history of the central 
power source of a quasar, which has been through subsequent phases of activity 
and quiescence. An exemplary source of that kind, quasar FIRST J164311.3+315618, 
was studied in \citet{Kunert2011}, 
and found to exhibit multiple radio structures.
 Another class of objects, which are claimed to contain the BH accretion disk, are
the Ultraluminous X-ray sources (ULXs). ULXs are a class of sources that have a 
luminosity larger than the Eddington one for the heaviest stellar-mass
objects ($ > 10^{40}$ ergs s$^{-1}$). Therefore, ULXs are frequently claimed to contain
accreting black holes with masses larger than the most massive stars and 
lower than AGNs ($10^3 - 10^6 M_{\odot}$ intermediate-mass black holes, IMBHs). An example object in this class is
HLX-1, which is possibly the best known candidate for an IMBH \citep{Farrell2009,Lasota2011HLX,Servillat2011,Godet2012}. This source is
located near the spiral galaxy ESO 243-49 \citep{Wiersema2010HLX,Soria2013} with peak luminosity exceeding $10^{42}$ erg s$^{-1}$. HLX-1 also exhibits periodic
limit-cycle oscillations. During seven years of
observations of its X-Ray variability, six significant bursts lasting a few tens of days
have been noticed. The mass of the black hole inside HLX-1 is estimated at about $10^4 - 10^5  M_{\odot}$ \citep{Straub2014HLX}. 
 In this work, we investigate a broad range of theoretical models 
 of radiation-pressure-driven flares and prepare the results
for easy confrontation with observational data. The appearance of the radiation pressure instability in hot 
parts of accretion disks can lead to significant 
outburst for all scales of the black hole mass. Temperature and heat production
 rate determine the outburst frequency and shape.
Different effective prescriptions for turbulent viscosity 
affect the instability
range and the outburst properties. Thus, by confronting the model 
predictions with observed flares, we can put constraints on those built-in assumptions.
The attractiveness of the radiation pressure instability as a mechanism to explain various phenomena across the whole black hole mass scale has been outlined by \citet{Wu2016}. In the present paper we expand this work by making a systematic study showing how the model parameters modify the local stability curve and global disk behaviour. We expand the parameter space of the model and identify the key parameters characterising the data. We develop a convenient way to compare the models to the data by providing simple fits to model predictions for the parameters directly measurable from the data. This approach makes the dependence of the models on the parameters much more clear, and it allows for much easier comparison of the model with observational data. 
\section{Model}
\subsection{Radiation pressure instability}
\label{sect:models}
 In the $\alpha$-model of the accretion disk, the non-zero 
component $\tau_{r \phi}$
of the stress tensor is assumed to be proportional to the total pressure.
The latter includes the radiation pressure component, which scales 
with temperature as $T^{4}$ and blows up in hot, optically thick disks for large accretion rates.
In general, an adopted assumption about the dependence of the $\tau_{r \phi}$ on the 
local disk properties leads to a specific prediction of the behaviour of the disk heating.
This in turn affects the heating and cooling balance between the energy 
dissipation and radiative losses.
Such a balance, under the assumption of hydrostatic equilibrium, is
calculated numerically with a closing equation for the
locally dissipated flux of energy given by the black hole mass and 
global accretion rate.
 The local solutions of the thermal balance and structure of a stationary accretion disk
at a given radius may be conveniently plotted 
in the form of a so-called stability 
curve of the shape $S$. 
Here, distinct points represent the annulus in a disk,  with temperature and surface density determined by the
accretion rate. For small accretion rates, the disk is gas pressure dominated and stable.
The larger the global accretion rate, the more annuli of the disk will be
affected by the radiation pressure and the extension of the instability zone grows 
in radius. The hottest areas of the disk are heated rapidly, the density 
decreases, and the local accretion rate grows; more material is 
transported inwards. The disk annulus empties because of both increasing 
accretion rate and decreasing density, so there is no self-regulation 
of the disk structure.
 However, the so called `slim-disk' solution \citep{1988ApJ...332..646A}, 
where the advection of energy
provides an additional source of cooling in the highest accretion rate regime
(close to the Eddington limit), provides a stabilising branch. Hence, 
the advection of some part of energy allows the disk to survive and oscillate 
between the hot and cold states. Such oscillating behaviour leads to periodic 
changes in the disk luminosity.
To model such oscillations, obviously, 
the time-dependent structure of the accretion
flow needs to be computed, instead of the stationary solutions described by the $S$-curve.
\subsection{Equations}
We are solving the time-dependent vertically averaged disk equations with radiation pressure in the Newtonian approximation,
 following the methods described in \citet{2002ApJ...576..908J}, and subsequent papers \citep{2005MNRAS.356..205J, Czerny2009,Janiuk2015}.
 The disk 
is rotating around the central object with mass $M$
with Keplerian angular velocity $\Omega = \sqrt{\frac{GM}{r^3}}$,
and maintains the local hydrostatic equilibrium in the vertical direction $P = C_3 \Sigma \Omega^2 H$. The latter gives
 the necessary relationship between pressure $P$, angular velocity $\Omega,$ and disk vertical thickness $H$. 
 $C_3$ is the correction factor regarding the vertical structure of the disk. In our model
 $ C_3 = \frac{ \int_{0}^{H} dz P(z)}{P(z=0) H}$.
 The vertically averaged stationary model is described in the paper \citep{2002ApJ...576..908J}.
 In the stationary (initial condition) solution, the disk is emitting the radiation flux proportional to the accretion  rate,
$\dot{M}$,  fixed by  the mass
and energy conservation laws 
\begin{equation}
 F_{\rm tot} = Q_+ = \frac{3 G M \dot{M}}{8 \pi r^3} f(r),
 \label{eq:ftot}
\end{equation}
where $f(r) = 1 - \sqrt{r_{in}/r}$ is the inner boundary condition term. 
 In the time-dependent solution the accretion rate is a function of radius, and the assumed accretion rate 
at the outer disk radius forms a boundary condition.
 We solve two partial differential equations describing the viscous and thermal evolution of the disk:
\begin{equation}
\frac{\partial \Xi }{\partial t}  = \frac{12}{y^2} \frac{\partial^2 }{\partial y^2} \Bigg[ \Xi \nu \Bigg], 
\label{eq:diffusion}
\end{equation}
where $y = 2 r^{1/2}$ and $\Xi = 2 r^{1/2} \Sigma$, and
\begin{equation}
\frac{\partial \ln T}{\partial t} + v_r \frac{\partial \ln T}{\partial r} = q_{\rm adv} + \frac{Q_{+} - Q_{-}}{(12-10.5 \beta)PH}.
\label{eq:energy}
\end{equation}
 The first equation represents the thin accretion disk's mass diffusion, and the second equation is the energy conservation. 
Here $\nu$ is the effective kinematic viscosity coefficient, connected with stress tensor as follows:
\begin{equation}
T_{r \phi} = \rho \nu r \frac{\partial \Omega}{\partial r}.
\end{equation}
 Thus, calculation of $\nu$ requires the assumption about the heating term.
The radial velocity in the flow is given by:
\begin{equation}
v_r = - \frac{3}{\Sigma} r^{-1/2} \frac{\partial}{\partial r} \Bigg[ \nu \Sigma r^{1/2} \Bigg].
\end{equation}
 The quantity $\beta$ is the ratio between gas and total (gas and radiation) pressure)
$\beta=\frac{P_{\rm gas}}{P}$. 
The advection term,
\begin{equation}
q_{\rm adv} = \frac{4 - 3 \beta}{12 - 10.5 \beta}  \Bigg[ \frac{d \ln \rho}{dt} + v_r \frac{d \ln \Sigma}{d r} \Bigg],
\end{equation}
is computed through the radial derivatives. For the initial condition $q_{\rm adv}$ is taken to be a constant, of order unity. 
The vertically averaged heating rate is given by:
\begin{equation}
Q_{+} = C_1 \tau_{\rm r \phi} r \frac{\partial \Omega}{\partial r} H,
\label{eq:qplus}
\end{equation}
where $\tau_{r \phi}$ 
is the vertically averaged stress tensor:
\begin{equation}
 \tau_{\rm r \phi} = \frac{1}{2H} \int_{-H}^{H} dz    T_{r \phi},
\end{equation}
and the radiative cooling rate per unit time per surface unit is
\begin{equation}
Q_{-} = C_2 \frac{4 \sigma_{\rm B} T^{4}}{3 \kappa \Sigma}.
\label{eq:qminus}
\end{equation}
where $\kappa$ is the electron scattering opacity, equal to 0.34  cm$^{-2}$ g$^{-1}$.
Coefficients $C_1$ and $C_2$ in Eqs. $(\ref{eq:qplus})$ and $(\ref{eq:qminus})$ are 
derived from the averaged stationary disk model \citep{2002ApJ...576..908J} and are equal
to: $C_1 = \frac{\int_0^H dz \rho(z)}{\rho(z=0) H } = 1.25$ , $C_2 =  6.25$ respectively.
\subsection{Expression for the stress tensor - different prescriptions}
The difficulty in finding the proper physical description of the turbulent behaviour of gas in the
ionised area of an accretion disk led to the adoption of several distinct theoretical prescriptions of
the non-diagonal terms in the stress tensor  term $\tau_{\rm r \phi}$.
Gas ionisation
should lead to the existence of a magnetic field created by the moving 
electrons and ions. 
The magnetic field in the disk is turbulent and remains in 
thermodynamical equilibrium with the gas in the disk. For a proper description
of the disk viscosity, different complex 
phenomena should be included in $ \tau_{r \phi}$. 
 For a purely turbulent plasma we can expect the  proportionality between the density of  kinetic energy of the 
gas particles and the energy of the magnetic field \citep{1973A&A....24..337S}. However, the disk   
geometry allows the magnetic field energy to escape \citep{1989ApJ...342...49S,2000ApJ...535..798N}.
Following \citet{1973A&A....24..337S}, one can therefore assume 
the non-diagonal terms in the 
stress tensor are proportional to the total pressure with a 
constant viscosity $\alpha$:
\begin{equation}
 \tau_{\rm r \phi} = \alpha P.
 \label{eq:shakura}
\end{equation}
 
On the other hand, \citet{1974ApJ...187L...1L} proved the instability of the model
described by \citet{1973A&A....24..337S}. Following that work,
 \citet{Sakimoto} proposed another 
formula which led to a set of stable solutions without any appearance of 
the radiation pressure
instability:
 \begin{equation}
  \tau_{\rm r \phi} = \alpha P_{\rm gas}.
   \label{eq:lightman}
\end{equation}
Later, Merloni and Nayakshin \citep{2006MNRAS.372..728M},
 motivated by the heartbeat states of GRS 1915+105, investigated the square-root formula
\begin{equation}
  \tau_{\rm r \phi} = \alpha \sqrt{P P_{\rm gas}}.
    \label{eq:merloni}
\end{equation}
This prescription was introduced by \citet{TaamLin1984} in the context of the Rapid Burster and used later by
 \citet{DoneDavis2008} and \citet{Czerny2009} both for galactic sources and AGNs.

In the current work, we apply a more general approach 
and introduce the entire family of models, with the contribution of the radiation pressure to the stress tensor 
parameterised by a power-law relation with an index $\mu \in [0,1]$.
We therefore construct a continuous transition between the disk, which is 
totally gas pressure dominated, and  the radiation pressure that 
influences 
the heat production \citep{1990MNRAS.244..377S,Honma1991,Watarai2003,2006MNRAS.372..728M}.
The formula for the stress tensor is a generalisation of 
the formulae in Equations $\ref{eq:shakura},\ref{eq:lightman}, \ref{eq:merloni}$ and is given by:
\begin{equation}
   \tau_{\rm r \phi} = \alpha {P^\mu P_{\rm gas}^{1 - \mu}}.
    \label{eq:szusz}
\end{equation}
In this work, we investigate the behaviour of the accretion disk 
described by formula  ($\ref{eq:szusz}$) 
for a very broad range of black holes and different values of $\mu$. 
A similar analysis has been performed also by \citet{2006MNRAS.372..728M} for different values of 
$\alpha$ (here, we fix our $\alpha$ with a constant value, which is 
at the level of $0.02$). 
Regarding the existence of a magnetic field inside the accretion disk, the viscosity can be magnetic in origin, and can reach different values 
for differently magnetised disks. As the strong global magnetic field can stabilise the disk \citep{Czerny2003,Sadowski2016}, the parameter 
$\mu$ can be treated as an effective 
prescription of magnetic field. 
\subsection{Numerics}
We use the code {\it GLADIS} (GLobal Accretion Disk InStability),
whose basic framework was initially described by 
\citet{2002ApJ...576..908J}. 
The code was subsequently developed 
and applied in a number of works to model the evolution of accretion disks
in Galactic X-ray binaries and AGNs
\citep{2005MNRAS.356..205J,Czerny2009,2012A&A...540A.114J}.
The code allows for computations with a variable time-step down to 
a thermal timescale, adjusting to the speed of local changes of the disk structure. 
Our method was recently used for modelling the behaviour of the microquasar
IGR J17091 \citep{Janiuk2015}. In that work, we used 
the prescription for a radiation-pressure dominated disk 
(with $\mu=1$, explicitly), but with an explicit formula approximating wind outflow,
which regulates the amplitudes of the flares, or even temporarily leads to a completely stable disk.
Here, we modified the methodology, and added the parameter $\mu$, 
which allows for a continuous transition between the gas and radiation pressure dominated cases, for example, with $\mu \in [0,1]$, as described by Equation \ref{eq:szusz}. We neglect the wind outflow, though, as in many sources the observable constraints for its presence are too weak. 
\subsection{Parameters and characteristics of the results}
 We start time-dependent computations from a certain initial state. This evolves for some time until the disk develops a
specific regular behaviour pattern. 
 We can get either a constant luminosity of the disk (i.e. stable solution), flickering, or 
periodic lightcurves, depending on the model parameters.
We parameterise the models by the global parameters: the black hole mass $M$, 
the external accretion rate, as well
as the physical prescription for the stress, $\alpha$ and $\mu$. 
 In this study we  mostly limited our modelling to a constant (arbitrary) value of $\alpha = 0.02$, since the 
scaling with $\alpha$ is relatively simple, and we wanted to avoid computing the four-dimensional grid of the models. We discuss our motivation below, and we also perform a limited analysis of the expected dependence on this parameter.

 Those parameters are not directly measured for the observed sources. In the current work, we focus on the unstable accretion disks. We thus construct from our models a set of output parameters that can be relatively easily measured from the observational data: the average bolometric luminosity $L$, the maximum bolometric luminosity $L_{\rm max}$, the 
minimum bolometric luminosity $L_{\rm min}$, the relative amplitude of a flare, 
$A = \frac{L_{\rm max}}{L_{\rm min}}$), and the period, $P$.

 In order to parameterise the shape of the light curve, we also introduce a dimensionless 
parameter $\Delta$, which is equal to the radio of the 
flare duration to the period. 
We define flare duration as the
time between the moments where the luminosity is equal to
$(L_{\rm max} + L_{\rm min})/2$ on the ascending slope of the flare, and the luminosity
$(L_{\rm max} + L_{\rm min})/2$ on the descending slope of the flare. 
We then compare $L$, $A$, $P$
and $\Delta$ obtained for several distinct black hole mass scales. 
In Table \ref{tab:input} we summarise 
the model input parameters. The accretion rate $\dot{m}$ is presented in Eddington units
$\dot{m} = \frac{\dot{M}}{\dot{M}_{Edd}}$. The Eddington accretion rate $\dot{M}_{Edd}$ is
proportional to the Eddington luminosity, and inversely proportional to the accretion efficiency
$\dot{M}_{Edd} = \frac{L_{Edd}}{\eta c^2}$. In this work, we focus on the case of an accretion disk
around a Schwarzschild BH with radius $R = 6 \frac{GM}{c^2}$. We assume accretion efficiency 
$\eta = \frac{1}{16}$ for all models in this paper.

In Table \ref{tab:res} we summarise the probed 
characteristics of the resulting flares.
 
  \begin{table}
  \begin{tabular}{|c|c|}
  \hline
   {\bf Value} & {\bf Symbol}
   \\ \hline
   Black hole mass &  $M$
     \\ \hline
  Accretion rate & $\dot{m}$
  \\ \hline
  First viscosity parameter  & $\alpha$
   \\ \hline
 Second viscosity parameter  & $\mu$
  \\ \hline
  \end{tabular}
\caption{Summary of the model input parameters }
\label{tab:input}
\end{table}

\begin{table}
  \begin{tabular}{|c|c|}
  \hline
   {\bf Value} & {\bf Symbol}
   \\ \hline
   Bolometric luminosity &  $L$
     \\ \hline
  Period & $P$
  \\ \hline
  Amplitude ($L_{\rm max}/L_{\rm min}$) & $A$ 
   \\ \hline
 Flare duration to period ratio  & $\Delta$
  \\ \hline
  \end{tabular}
\caption{Summary of the characteristic quantities used to describe the
 accretion disk flares}
\label{tab:res}
\end{table} 
\subsubsection{Value of $\alpha$ viscosity parameter}
\label{sect:valalpha}
Our choice of $\alpha = 0.02$ as a reference value was motivated by observations of AGNs.  \citet{SiemiginowskaCzerny1989} interpreted the quasar variability as the local thermal timescale at a radius corresponding 
to the observed wavelength in the accretion disk, and they determined the value of 
$0.1$ for a small sample of quasars. The same method, for larger sample objects 
(Palomar-Green quasar sample), gave the constraints $0.01 < \alpha < 0.03$ for sources with luminosities
$0.01 L_{Edd}< L < L_{Edd}$ \citep{Starling2004}. Values in the range $0.104 < \alpha < 0.337$ were
found for blazars from their intra-day variability \citep{Xie2009} but those variations, even if related
to the accretion disk, might be coming from Doppler-boosted emission and the timescale
is then under-estimated.
 The stochastic model of AGN variability \citep{Kelly2009,Kozlowski2016} allows for determination of the characteristic
timescales and 
their scaling. \citet{Kelly2009} give the value of the viscosity parameter $\alpha = 10^{-3}$ estimated at the
distance of 100 $R_{Schw}$ , but the actual value implied by Eq. (5) { in \citet{Kelly2009}}  is 0.2. This value would be lower if the radius was smaller. More precise results come from \citet{Kozlowski2016}. From this paper we have
\begin{equation}
\tau_{\rm char} [years] = 0.97 \Bigg[ {M \over 8 \times 10^8 M_{\odot}} \Bigg] ^{0.38 \pm 0.15}.
\end{equation}
This characteristic timescale is obtained at a fixed wavelength band, or disk temperature, and the location of a fixed
disk temperature $T$ in the Shakura-Sunyaev disk also depends on the black hole mass.
We thus identify this timescale with the thermal timescale, and obtain an expression for the viscosity 
parameter
\begin{equation}
\alpha = 0.4 ({T \over 10^{4} K})^{-2} \Bigg[ {M \over 8 \times 10^8 M_{\odot}} \Bigg] ^{0.12 \pm 0.15}.
\end{equation}
We see that the viscosity does not depend on the black hole mass within the error. The variability study 
of \citep{Kozlowski2016} was perfomed predominantly in the $r$ SDSS band (6231 \AA), quasars being mostly at
redshift 2. The conversion between the local disk temperature and the maximum disk contribution at a given wavelength is given approximately as $ h \nu = 2.43 k T$ (where $h$ and $k$ are the Planck and Boltzmann constants).
Therefore the dominant temperature in  the \citet{Kozlowski2016} sample is about 28 000 K, and the corresponding viscosity
parameter is 0.044 for the black hole mass $\sim 8.0 \times 10^8 M_{\odot}$ and 0.015 for $\sim 8.0 \times 10^4 M_{\odot}$. 
In following sections of this paper, we  thus fix the parameter $\alpha = 0.02$ in most of the models. For the exact determination of the HLX-1
mass we tried to slightly change this parameter. For the case of microquasars, we computed a small grid for one particular value of 
mass, accretion rate, and $\mu$. We noticed the power dependencies between the observables (P, A, $\Delta$) and the value of $\alpha$. 
These functions are described  in Sect. \ref{sect:ma}.
\section{Local stability analysis}
\label{sect: stability}
We first perform a local stability analysis in order to formulate basic
expectations and limit the parameter space. In general, the disk
is locally thermally unstable if for some temperature $T$ and radius $r$,
the local heating
rate grows with the temperature faster than the local cooling rate:
\begin{equation}
 \frac{d \log Q_+}{d \log T} > \frac{d \log Q_-}{d \log T}.
 \label{eq:instcondition}
\end{equation}
In this analysis, we consider timescales that are different for the thermal
and viscous phenomena which is justified for a thin disk.
\subsection{Stability and timescales}
For thin, opaque accretion disks, we have strong timescale separation between thermal (connected with the local heating and cooling)
and viscous phenomena. The thermal timescale, for a disk rotating with angular velocity $\Omega$, is $t_{\rm th} = \alpha^{-1} \Omega^{-1}$.
The appearance of viscous phenomena, connected with the large-scale angular momentum transfer is connected with the disk thickness, so that $t_{\rm visc} = t_{\rm th} \frac{R^2}{H^2}$.
We focus now on the thermal phenomena. On thermal timescales the 
local disk surface density is constant, and only the vertical inflation is allowed. Therefore, for that timescale we can assume $\Sigma = const$.
From Eqs. $(\ref{eq:qplus})$ and $(\ref{eq:merloni})$ we have:
\begin{equation}
 Q+ = \frac{3}{2} C_{1} \alpha P^\mu P_{\rm gas}^{1- \mu} H \Omega.
 \label{eq:qplus2a}
\end{equation}
Assuming that the disk maintains the vertical hydrostatic equilibrium $H = P/( C_3 \Sigma \Omega^2)$,
and defining $x = \frac{P_{\rm gas}}{P_{\rm rad}} + 0.5$,
 we can rewrite Eq. $(\ref{eq:qplus2a})$ as:
\begin{equation}
 Q_+ = \alpha \frac{3}{2 C_3 \Sigma \Omega^2} P^{1+\mu} P_{\rm gas}^{1- \mu} =  \frac{3}{ 2} \frac{\alpha}{C_3 \Sigma \Omega^2} P_{\rm rad}^2 (x + 1/2)^{1+\mu}(x - 1/2)^{1-\mu}.
 \label{eq:qplus2}
\end{equation}
Then, if we assume a constant $\Sigma$ regime, we have:
\begin{equation}
 \frac{d x}{d T} = - \frac{7}{2} \frac{(x+ 1/2)(x-1/2)}{2xT}
\end{equation}
and
\begin{equation}
 \frac{d \log Q_+  }{ d \log T } = 1 + 7 \mu \frac{1 - \beta}{1 + \beta},
 \end{equation}
 where $\beta = \frac{P_{gas}}{P} = \frac{x - 1/2}{x+ 1/2}$. 
 Finally, from Eq. (\ref{eq:qminus}) and  Eq. (\ref{eq:instcondition}), we have:
 \begin{equation}
  \frac{d \log Q_+  }{ d \log T } > 4,
 \end{equation}
which is fulfilled if the condition: 
 \begin{equation}
  \beta < \frac{7 \mu - 3}{7 \mu + 3} 
    \label{eq:muszusz}
 \end{equation}
is satisfied \citep{1990MNRAS.244..377S}.
 This gives the necessary condition for the instability for the case of $\mu$-model, so that the instability occurs
only if $\mu > 3/7$.
\subsection{Magnetised disk and its equivalence to $\mu$ model}
The existence of strong magnetic fields can stabilise the radiation-pressure dominated disk \citep{ZdziarskiSvensson1994,Czerny2003,Sadowski2016}. 
We can assume a significant magnetic contribution to the total pressure 
$P$, defining it as follows:
\begin{equation}
 P = P_{\rm rad}+P_{\rm gas}+P_{\rm mag}.
 \label{eq:pmag}
\end{equation}
Let us define the disk magnetisation coefficient 
$\beta' = \frac{P_{mag}}{P}$. We put the formula $(\ref{eq:pmag})$
into the Shakura-Sunyaev stress-energy tensor (i.e. for 
$\mu = 0$ in Eq. $\ref{eq:qplus2}$), and then we get:
\begin{equation}
   \frac{d \log Q_+  }{ d \log T }  = 8(1 - \beta').
\end{equation}
Here, the value of $\beta' = \frac{1}{2}$ means that there is an
 equipartition between the energy density of the 
gas radiation and magnetic energy density. It corresponds to the complete
stabilisation of the disk, so that $ \frac{d \log Q_+  }{ d \log T } \le  \frac{d \log Q_-  }{ d \log T }$.
From the formula  $(\ref{eq:muszusz})$ we can connect $\beta'$ and $\mu$ as follows:
\begin{equation}
 \mu = 1 - \frac{8}{7} \beta'.
 \label{eq:mumagnetic}
\end{equation}
Regarding the observed features, the model of the magnetised disk is equivalent to the $\mu$ 
model for the radiation-pressure dominated disks in terms of appearance of thermal
instability. The major difference is that the greater total pressure makes 
the disk 
thicker. This fact can have some influence on the behaviour of the light curve,
which we discuss below.
\subsection{Disk magnetisation and amplitude}
The energy equation $(\ref{eq:energy})$ can give us a direct connection between the heating rate and pressure.
If we assume that the heating rate dominates the cooling rate, we have: 
\begin{equation}
 \frac{d T}{d t} = C T^{7 \mu - 7},
\end{equation}
where $C$ is a constant on the thermal timescale and depends on the 
local values of $\Sigma$ and
$\Omega$.
This simple, first-order differential equation 
 gives us the following dependence on the heating growth:
\begin{equation}
 \frac{d \log T}{d \log t} = \frac{1}{7(1 - \mu)}.
 \label{eq:ode}
\end{equation}
It explains why flares are sharper for bigger $\mu$. It can also 
give another criterion for determination of a proper value of $\mu$, and
can be used as a test for the validity of $\mu$-model in confrontation with the observational data. If we assume that
most of the luminosity comes from a hot, thermally unstable region of the disk  \citep{Janiuk2015} and we apply it to $(\ref{eq:ode}) $, we get:
\begin{equation}
 \frac{d \log L}{d \log t} = \frac{4}{7(1 - \mu)}.
\end{equation}
Regarding the timescale separation, and assuming that the flaring
of the disk
is stopped by the viscous phenomena after $t \approx t_{\rm visc}$,
we get the following formula for the dependence between the 
relative amplitude $A = \frac{L_{\rm max}}{L_{\rm min}}$ and the viscous to thermal timescales rate:
\begin{equation}
 \log A = \frac{4}{7} \frac{1}{1-\mu} \log \Bigg[ \frac{t_{\rm visc}}{t_{\rm th}} \Bigg].
\end{equation}
Furthermore, we can derive a useful formula that connects the 
presence of magnetic fields with the amplitude of the 
limit-cycle oscillations:
\begin{equation}
 \log A = \frac{1}{2} \frac{1}{ \beta'} \log \Bigg[ \frac{t_{\rm visc}}{t_{\rm th}} \Bigg].
\end{equation}
In the remainder of this article, we perform a more detailed analysis of the relation between the outburst amplitude and
other light curve properties on the $\mu$ parameter, which is 
possibly corresponding to the scale of magnetic fields.
 
 
\section{Results - stationary model}
\label{sect:resultsstationary}
\begin{figure}
\includegraphics[width=\columnwidth]{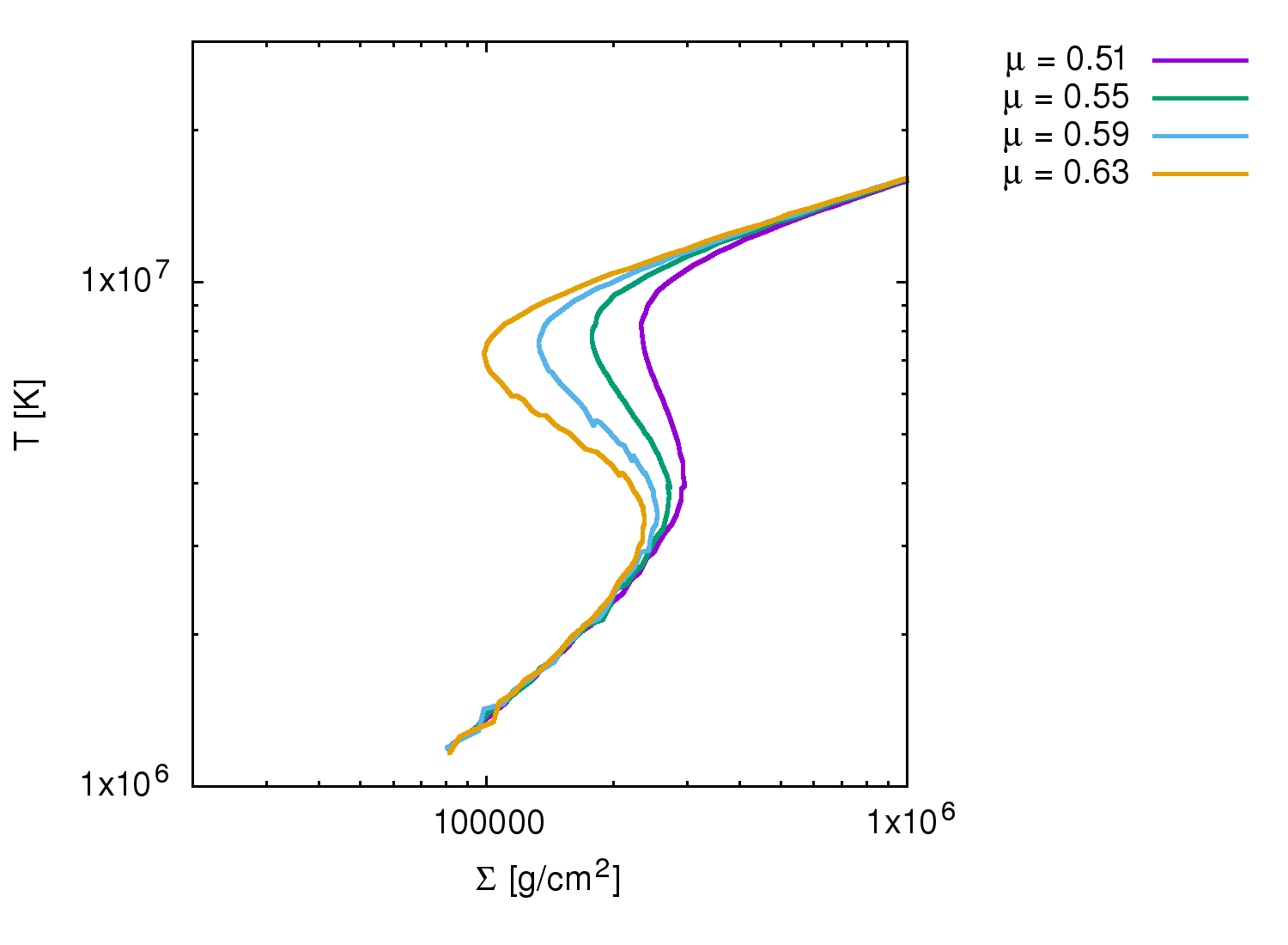}
\caption{Local stability curves for $\mu = 0.51$, $\mu = 0.55$, 
$\mu = 0.59$ and $\mu = 0.63$.
Parameters: $M = 3 \times 10^4 M_{\odot}$, $\alpha = 0.02$, and $\dot{m} = 0.7$. The chosen 
radius is $R =  7.82 R_{Schw} = 15.74 \frac{G M}{c^2} = 6.88 \times 10^{10} $ cm, 
corresponding to the inner, hot area of the disk. The curves
depend strongly on the $\mu$ parameter, and larger $\mu$ provides 
a larger negative
derivative range corresponding to the unstable state. }
\label{fig:stability}
\end{figure}

First, we compute an exemplary set of stationary models, to verify the  expected parameter range of the instability.
Eq. ($\ref{eq:muszusz}$) gives the relation
between the maximum gas-to-total pressure ratio and the minimum value of the $\mu$ 
parameter. We perform numerical computations for an intermediate 
black hole mass of $3 \times 10^4 M_\odot$, and plot the stability curve at 
the radius  
$R =  7.82 R_{Schw} = 15.74 \frac{G M}{c^2} = 6.88 \times 10^{10} $cm.
The disk is locally unstable if the 
slope  $ \frac{\partial \Sigma}{\partial T}$ is negative (Eq. $(\ref{eq:instcondition})$). It gives the necessary, 
but not the sufficient condition for the global instability as for the appearance of
the significant flares, the area of the 
instability should be sufficiently large. 
We compute the S-curves (see Section \ref{sect:intro}), which are presented in
Figure \ref{fig:stability} for different values of $\mu$. 
The bigger $\mu$, the bigger the negative slope area on
the S-curve, and therefore the larger the range of the instability.
 However, only the hot, radiation-pressure-dominated area of the disk remains unstable, and 
for larger radii the S-curve bend moves towards the enormously large, super-Eddingtonian external accretion rate. 
In effect, for sufficiently large radii,
the disk remains stable. The same trend is valid 
also for stellar mass (microquasars) and supermassive (AGN) black holes \citep{Janiuk2011}.
\begin{figure}
\includegraphics[width=\columnwidth]{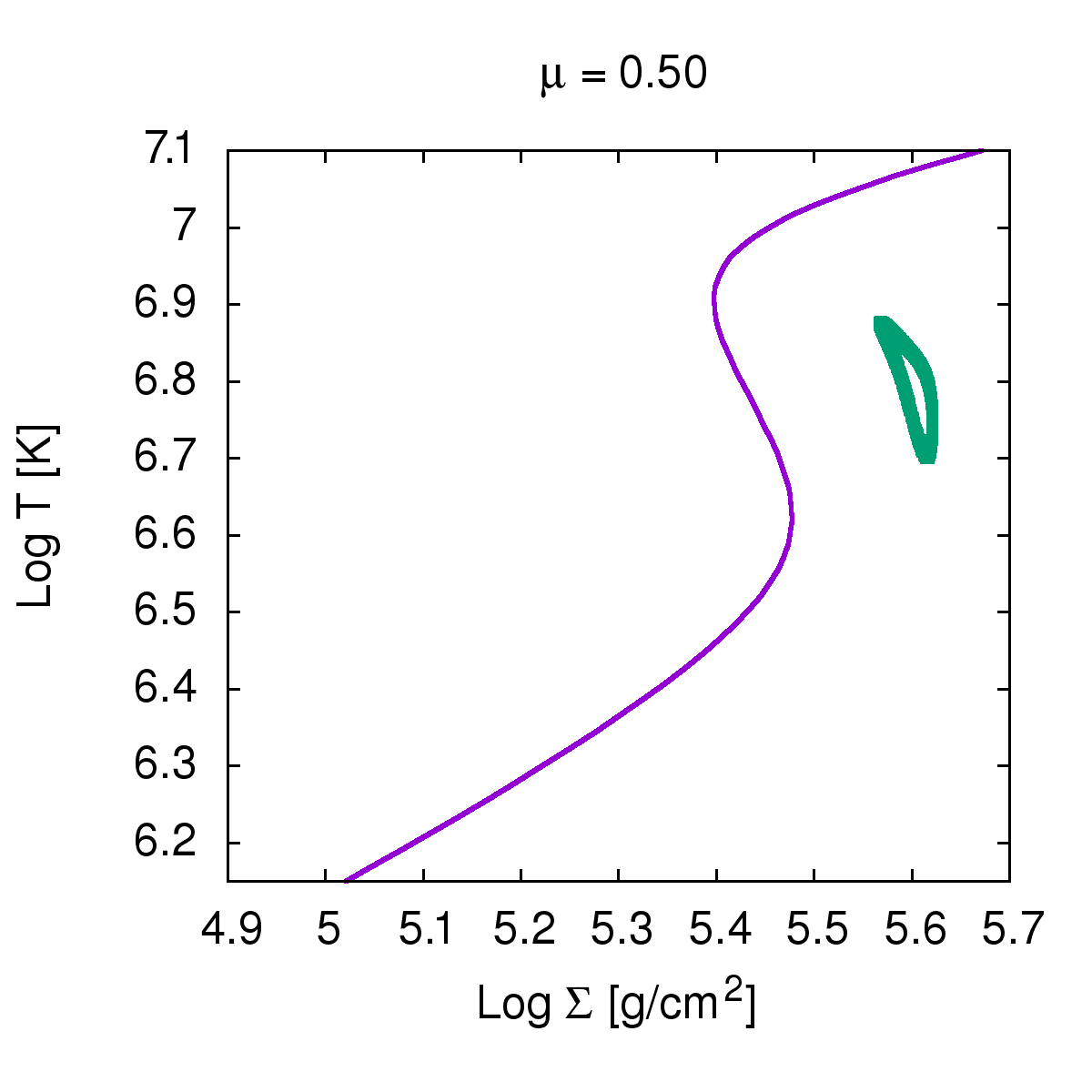}
\caption{$T$ and $\Sigma$ variability for the model with $\mu = 0.5$ 
for a typical IMBH accretion disk. The computation shows a
 weakly developed instability.
Parameters: $M = 3 \times 10^4 M_{\odot}$, $\alpha = 0.02$, and $\dot{m} = 0.25$. The plot is made for the
radius $R = 7.82 R_{Schw} = 15.74 \frac{G M}{c^2} = 6.88 \times 10^{10} $ cm. 
The red curve represents the stationary 
model and the green dots show local values of the temperature for the time-dependent computation.
}
\label{fig:cykl51}
\end{figure}
\begin{figure}
\includegraphics[width=\columnwidth]{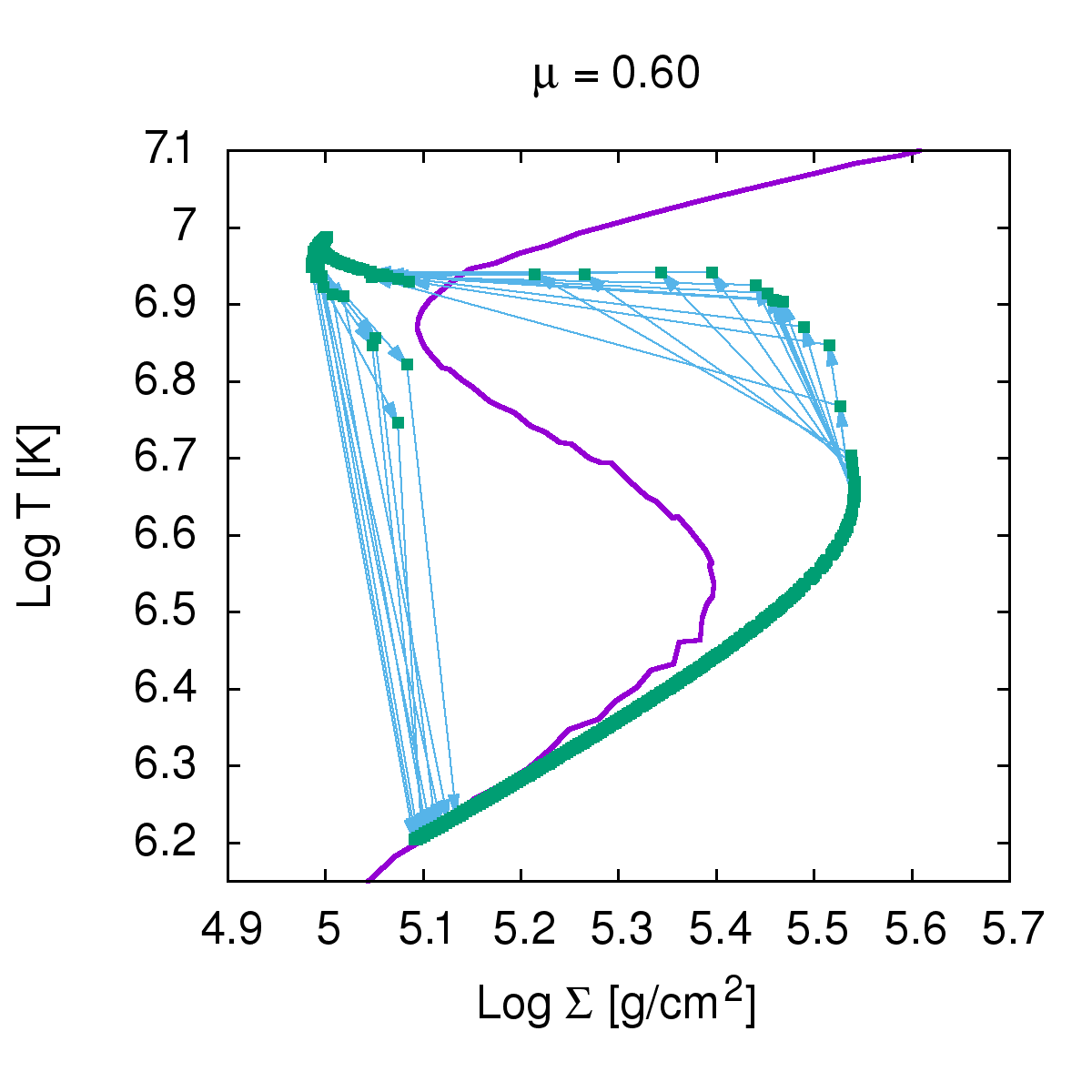}
\caption{$T$ and $\Sigma$ variability for the model with $\mu = 0.6$ for a typical IMBH accretion disk. 
The models present a strongly developed instability, 
leading to huge outbursts.
Parameters: $M = 3 \times 10^4 M_{\odot}$, $\alpha = 0.02$, and $\dot{m} = 0.25$ for the
radius $R = 7.82 R_{Schw} = 15.74 \frac{G M}{c^2} = 6.88 \times 10^{10} $cm. 
The red curve represents the stationary 
solution and the green dots show local values of the temperature and density 
in the dynamical model.
The points occupy both the upper and lower branches of the S-curve. Green points
represent the most common states of the disk, while blue vectors represent directions of the most common
rapid changes of local $T$ and $\Sigma$.}
\label{fig:cykl63}
\end{figure}

\section{Results - time dependent model}

\label{sect:results} 
 In this section, we focus on the numerical computations of the full time-dependent model. We perform the full computations of the radiation pressure instability models since the stability 
 curves give us only the information about the local disk stability. 
However, the viscous transport
 (Eq.(\ref{eq:diffusion})) and the radial temperature gradients deform
 the local disk structure, and the time evolution of the disk at a fixed radius resulting from the global simulations does not necessarily follow the expectations based on local stability analysis. 
 
   Figures \ref{fig:cykl51} and \ref{fig:cykl63} present the stability curves (red)
 and the global solutions of the dynamical model plotted at the same single radius (green). 
 Stronger bending of the shape of the light curves appears for larger $\mu$
 (Figs. \ref{fig:stability}, \ref{fig:cykl51}, and \ref{fig:cykl63}) resulting in
 broader development of the instability, visible in the shape of light curves
 (Figs. \ref{fig:ifl},\ref{fig:iou}). 
 Low values of $\mu$ cause the presence of the instability within a small range of 
radii, therefore the instability is additionally dumped by the stable zones. The time-dependent 
solution never sets on the stability curve, the covered area in Fig.\ref{fig:cykl51} is very small, 
and the corresponding global light curve shows only small flickering. 
The growth of the instability for  large values of $\mu$ results in 
the dynamical values of
$T$ forming two coherent sets (see Figure \ref{fig:cykl63}),
and the solution follows the lower branch of the stability curve. This part of the evolution describes the
prolonged period between the outbursts.

 For comparison 
 with the observed data we need to know the global time behaviour of the disk  for a broad parameter range.
 We ran a grid of models for typical stellar mass  $ M =  10 M_{\odot}$,
 for intermediate black hole mass $M = 3 \times 10^{4} M_{\odot} $, {and for 
 supermassive black holes ($M = 10^8 M_{\odot}$)}. We adopted $\alpha = 0.02$ throughout all the computations, and accretion rates ranged between $10^{-1.6}$ and $10^{-0.2}$ 
 in Eddington units.
 For each mass, we present the relations between 
 period, amplitude, and duration divided by period. 
 According to \citet{Czerny2009}, the threshold 
 accretion rate is (the sufficient rate for the flares to appear) is  
 at the level of 0.025 for AGNs. For our models, the critical value of accretion
 rate is at the level of
 $0.025 - 0.1$ of the Eddington rate. 
 Below, we present our results through a set of mutual correlations between
 observable characteristics of the flares, $P$, $A$, and $\Delta$ (see Table \ref{tab:input}), and the model parameters,
 $\dot{m}$, $M,$ and $\mu$. 
 \subsection{Light curve shapes}
\begin{figure}
\includegraphics[width=\columnwidth]{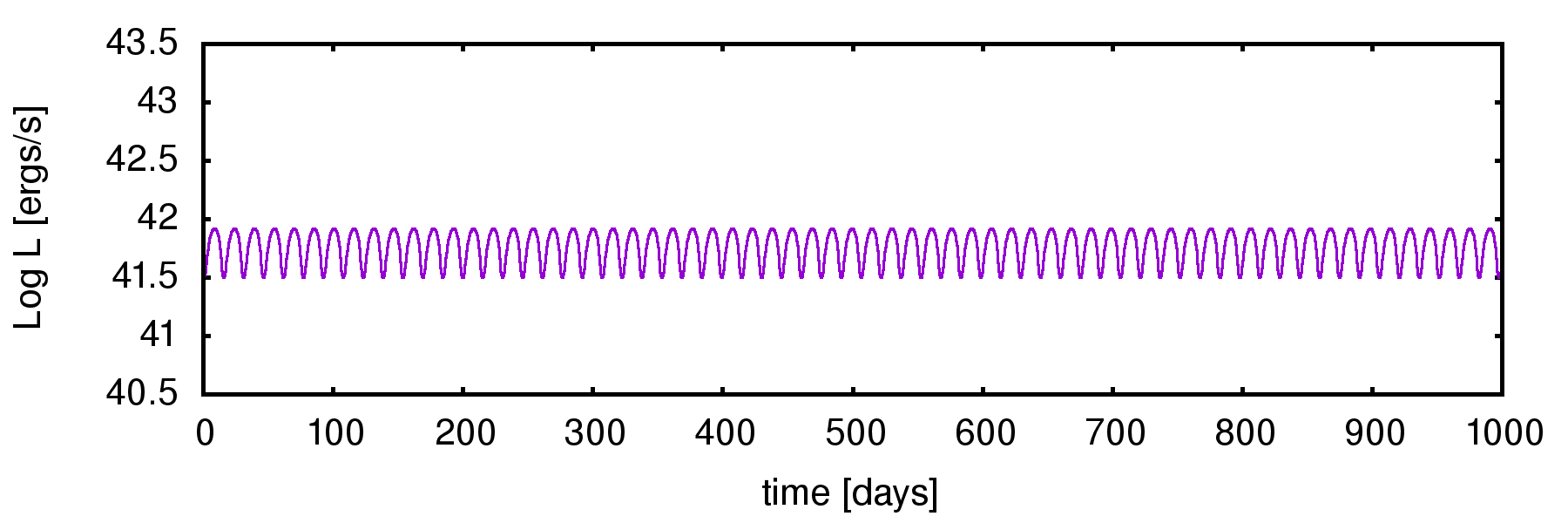}
\caption{Typical flickering light curve for intermediate mass black hole and smaller $\mu$ ($M = 3 \times 10^4$, $\mu = 0.5$ $\dot{m} = 0.25$)}
\label{fig:ifl}
\end{figure}

\begin{figure}
\includegraphics[width=\columnwidth]{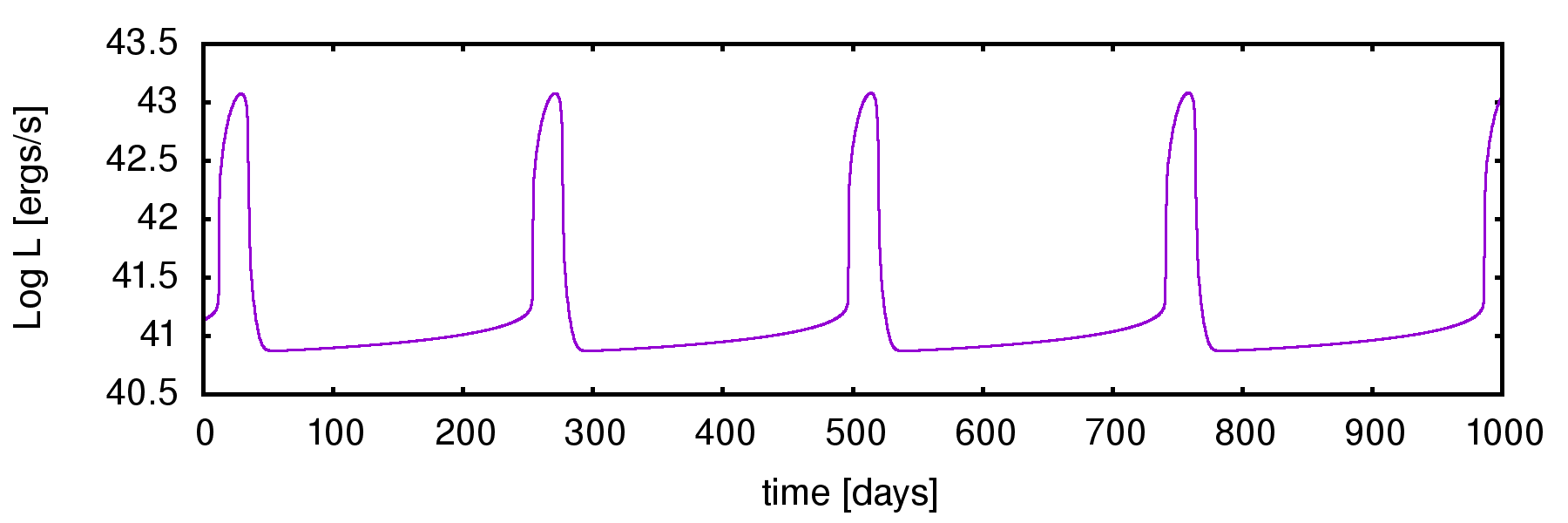}
\caption{Typical outburst light curve for intermediate mass black hole and larger $\mu$ ($M = 3 \times 10^4$, $\mu = 0.6$ $\dot{m} = 0.25$)}
\label{fig:iou}\end{figure}

 Here we define different characteristic modes of the flares. Since we have a dynamical system described by a set of
 non-linear 
 partial differential equations, we expect that the flares will form different patterns of variability, which should be comparable to 
 the observed patterns. For that non-linear system 
 we can distinguish between the flickering behaviour and the strong flares.
In Figures \ref{fig:ifl} and \ref{fig:iou}, we present typical cases of flickering and outburst flares. 
The difference
 between the flickering and outburst modes lies not only in their amplitudes;
as we can see in Figure \ref{fig:iou}, the long low luminosity 
 phase, when the inner disk area remains cool,
 is not present in the flickering case, presented in Fig. \ref{fig:ifl}.
The latter, corresponds to the temperature and density variations 
 presented in Fig. \ref{fig:cykl51},
 where the surface density of the disk does not change 
 significantly. In contrast,
 Fig. \ref{fig:iou} presents the case where 
 the surface density changes significantly and 
$\Sigma$ needs a long time to grow to the value where
rapid heating is possible. 
In the case of flickering
we can distinguish two phases of the cycle: (i) heating, when
 the temperature in the inner regions
of the disk is growing rapidly, and (ii) advective, when the inner annuli cool down significantly, and are then
extinguished when the disk is sufficiently cool. 
Now, after a strong decrease of advection, the heating 
phase repeats again.
In the case of the burst, the phases (i) and (ii) are much more
developed and advection is able to achieve instantaneous thermal equilibrium
of the disk, 
in contradiction to flickering, where the disk is always unstable.
The instantaneous equilibrium leads to the third phase (iii), diffusive, when the surface density in the inner annuli
of the disk is growing up to the moment when the stability curve has a negative slope, and the cooling rate, $Q_{-}$, is
significantly smaller in comparison to the heating rate, $Q_{+}$. Then, the phase of heating
repeats.

 \subsection{Amplitude maps}
 \label{maps}
\begin{figure}
\includegraphics[width=\columnwidth]{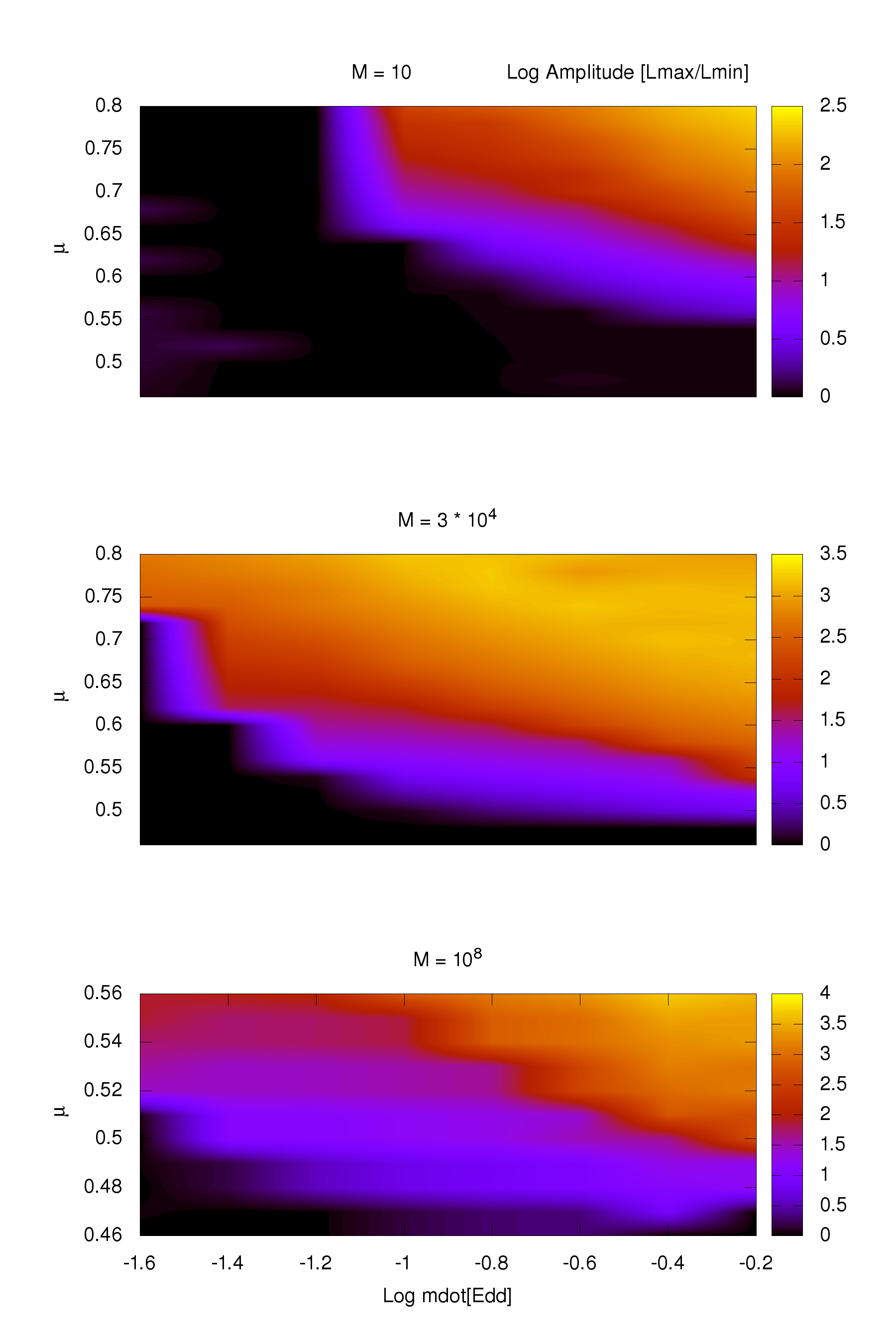}
\caption{ Amplitude map for different values of accretion rates and $\mu$ 
   for the accretion disk around a stellar-mass black hole ($M = 10 M_\odot$, upper panel),
  around an intermediate-mass black hole ($M = 3 \times 10^4 M_\odot$ , middle panel), and 
around a supermassive black hole ($M = 10^8 M_\odot$ , lower panel). 
Black regions   represent no flares and a lack of instability.}
 \label{fig:sa}
\label{fig:ia}
\label{fig:aa}
\end{figure}

  Figure {\ref{fig:sa}} show{s} dependencies between the accretion rate, $\mu$ coefficient, and flare amplitude. The black area corresponds to
  cases of stable solutions without periodic flares. 
  The violet area corresponds to a small flickering, and red and yellow 
  areas correspond to bright outbursts.
  Since for a given accretion rate and $\mu$ the AGN disks are much more radiation-pressure-dominated \citep{Janiuk2011}, the critical values of the accretion rates in
 Eddington units are the lowest for AGNs. Thus, the stabilising influence of a magnetic 
 field is more pronounced for the microquasars, than for AGNs.   
 Our results include different variability patterns. As shown in Figure
  \ref{fig:sa}, for a given set of $\mu$ and
  accretion rate, the relative amplitude varies by many orders of magnitude
  (from small flares, changing the luminosity by only a few percent, up to the 
large outbursts with 
  amplitudes of $\sim 200$ for microquasars, $\sim 1000$ for intermediate black holes, and $\sim 2000$
  for AGN). 
  The flare amplitude grows with accretion rate and with $\mu$.
  To preserve the average luminosity on sub-Eddington level, also the light curve
  shape should change with at least one of these parameters. 
  Let $\Delta$ be flare duration to period ratio, as defined
  in Section \ref{sect:intro}. 
  To preserve the average luminosity $L$, the energy emitted during the flare plus energy
  emitted during quiescence (between the flares) should be lower 
  than the energy emitted during one period. Since the radiation pressure instability reaches
  only the inner parts of the disk, the outer  stable parts of the disk  radiate 
  during the entire cycle, maintaining the luminosity at the level of $L_{\rm min}$. This level can be computed from Eq. $(\ref{eq:ftot})$
  since the outer border of the instability zone is known \citep{Janiuk2011, Janiuk2015}. 
 
\subsection{Amplitude and period}
\begin{figure}
\includegraphics[width=\columnwidth]{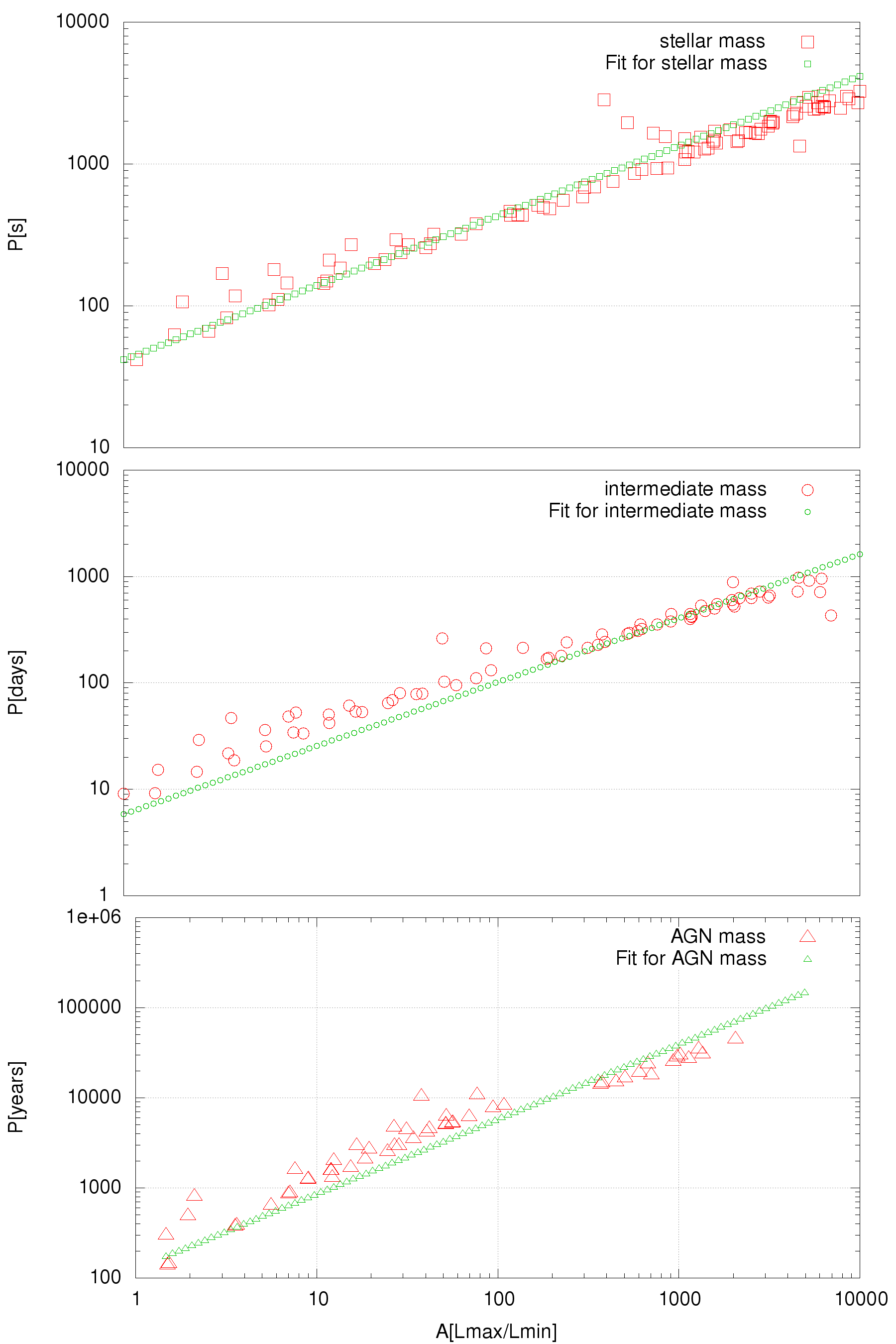}
\caption{ Dependence between period and amplitude of flares for a stellar-mass 
 black hole ($M = 10 M_{\odot}$, upper panel),  an 
intermediate-mass black hole ($M = 3 \times 10^{4} M_{\odot}$, middle panel), and  a supermassive 
black hole ($M = 10^{8} M_{\odot}$, lower panel). Computations were made for a range of $\mu$ but values for a different  $\mu$ lie predominantly along the main correlation trend resulting in a very low scatter.
 }
\label{fig:sap}
\label{fig:iap}
\label{fig:aap}
\end{figure}

  In Figure \ref{fig:sap}, we present the dependence between period and amplitude for the
 microquasars, intermediate mass black holes and AGN, respectively.
 In general, the amplitude grows with the period $P$, $\mu$, and accretion rate
 $\dot{m}$.
 { Figure \ref{fig:sap} was made for the three grids of these models which 
result in significant
 limit-cycle oscillations (i.e. $L_{\rm max}/L_{\rm min} > 1.2$). In our work we used regular and rectangular grids. 
 For $M = 10 M_\odot$ and $M = 3 \times 10^4 M_\odot$ we chose $\alpha = 0.02$ (as explained in 
 Sect. $\ref{sect:valalpha}$), and the parameter ranges were $\mu = \{ 0.44, 0.46, ... , 0.80 \}$ and 
 $\log \dot{m} = \{-1.6, ... -0.4, -0.2\}$. For $M = 10^8 M_\odot$ we chose a different
 range and sampling, $\mu = \{ 0.44, 0.45, ... , 0.56 \}$, but the same range of $\dot{m}$.
 The range of $\dot{m}$ was adjusted to cover all the values for which the instability appears.
The values of $\mu$ were chosen to consist of supercritical ones, according to Eq. (\ref{eq:muszusz}) (i.e. larger than $\mu_{\rm cr} = 3/7$). 
The upper cut-off of $\mu$
 (i.e. $0.8$ for XRBs and IMBHs, and $0.56$ for AGNs) was chosen because of 
computational complexity that arises for larger $\mu$. 
Nevertheless, the observed properties of the sources
(see, e.g. Table \ref{tab:deter} in Section \ref{sect:discussion}) suggest that in any case the values of $\mu$ are limited.
}

 The results shown by points in {Fig. \ref{fig:sap}} can be fitted with the following formula:
 \begin{equation}
  \log P ~ [{\rm sec}] \approx 0.83 \log \frac{ L_{\rm max}}{ L_{\rm min}} + 1.15 \log M  + 0.40.
  \label{eq:20160713}
 \end{equation}
 Here $P$ is the period in seconds and $M$ is the mass in Solar masses. The above general relation gains the following forms, if we want to use it for the sources
 with different black hole mass scales:
  \begin{equation}
  \log P_{\rm MICR} ~ [{\rm sec}] \approx 0.83 \log \frac{ L_{\rm max}}{ L_{\rm min}} + 1.15 \log \frac{M}{10 M_\odot}  + 1.55,\end{equation}
for the microquasars (see fit on Fig. \ref{fig:sap}), then
 \begin{equation}
  \log P_{\rm IMBH} ~ [{\rm days}]  \approx 0.83 \log \frac{ L_{\rm max}}{ L_{\rm min}}  + 1.15 \log \frac{M}{3 \times 10^4 M_\odot} + 0.53 
  \label{eq:20160713i}
 \end{equation}
for intermediate mass black holes (see fit on Fig. \ref{fig:iap}), and finally
  \begin{equation}
  \log P_{\rm AGN} ~ [{\rm years}] \approx 0.83 \log \frac{ L_{\rm max}}{ L_{\rm min}} + 1.15 \log \frac{M}{10^8 M_\odot} + 2.1
  \label{eq:20160713a}
 \end{equation}
for active galaxies (see fit on Fig. \ref{fig:aap}).

 From the formula (\ref{eq:20160713}) we can estimate the values of masses of 
 objects, if the values of $P$ and $A$ are known. The period-amplitude dependence is 
 universal, and, in the coarse approximation, does not depend on $\mu$. The positive
 correlation between period and amplitude indicates that those observables 
 originate in one nonlinear process, operating on a single timescale. Although the variability patterns can vary for 
 different accretion rates, for a given mass, the period and amplitude are strongly correlated
 and can describe the range of instability development. It can, in general, be adjusted by the specific model parameters, but the basic disk variability pattern is universal in that picture.
\subsection{Amplitude and width}
\label{sect:as}
\begin{figure}
\includegraphics[width=\columnwidth]{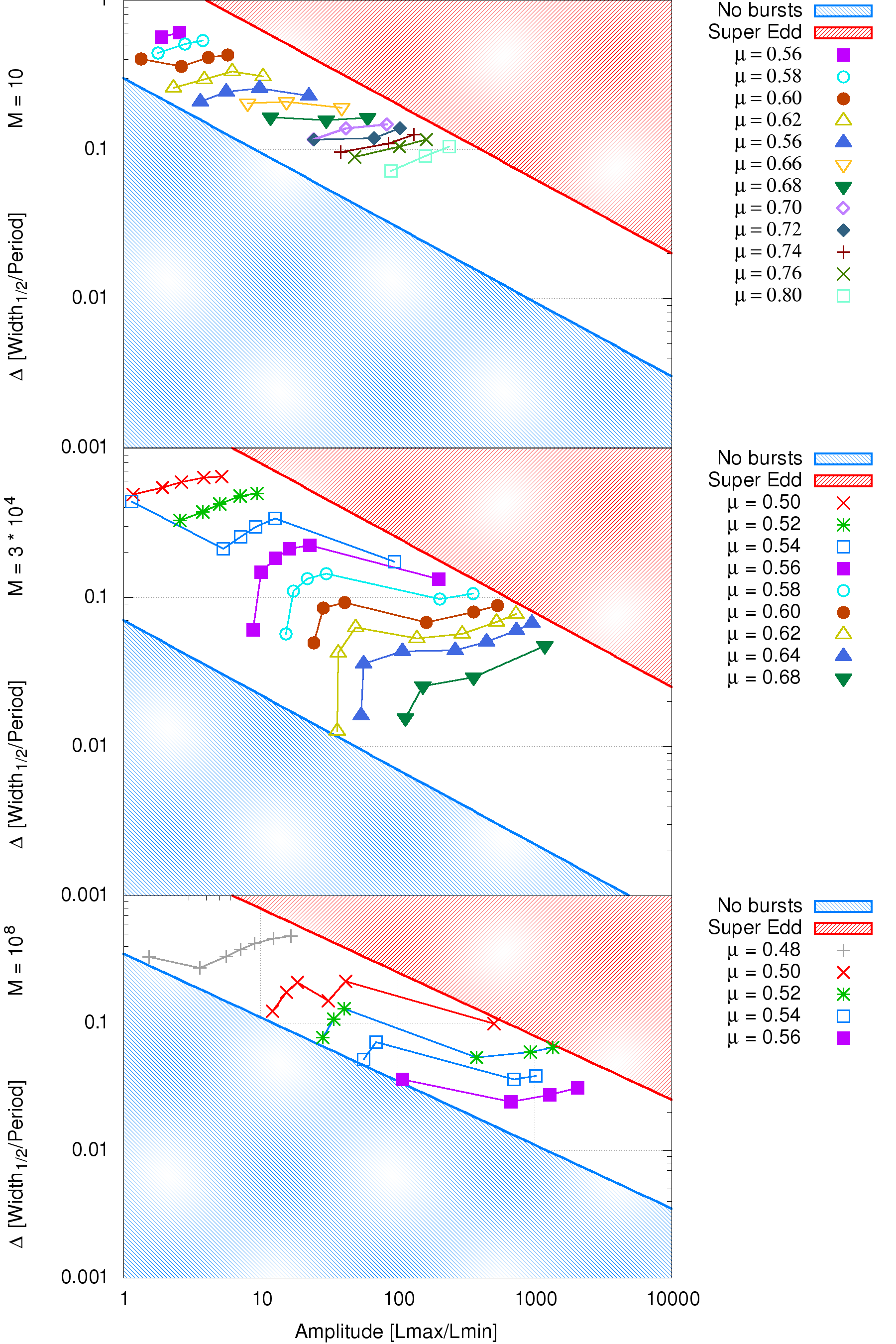}
\caption{ Dependence between the amplitude and width of flares for a stellar-mass 
black hole ($M = 10 M_\odot$, upper panel),  an intermediate-mass 
black hole ($M = 3 \times 10^4 M_\odot$, middle panel), and  a supermassive
black hole ($M = 10^8 M_\odot$, lower panel). The ranges of
$\mu$ are given for each figure. The colour lines represent isolines for different 
$\mu$. We note that the scatter is now larger than in Fig. \ref{fig:sap}. }
\label{fig:sas}
\label{fig:ias}
\label{fig:aas}
\label{fig:a}
 \end{figure}

In Figure \ref{fig:sas} the relation between the flare amplitude and its width is presented. 
Both values are dimensionless and show similar reciprocal behaviour for the black hole accretion disks flares
 across many orders of magnitude. The value of $\Delta$ can help us to distinguish
between different states of the source. It should be also noted that $\Delta$ depends
on the amplitude of the outburst only for small amplitudes, while for the 
larger ones, $\Delta$ remains approximately 
constant. The most convenient classification
 is to introduce the `flickering' mode (A < 50), which corresponds to the
 large ratio of the flare duration to its period
 ($\Delta > 0.15$),
 and the `outburst' mode (A > 50 ), which corresponds to the small flare duration to period ratio  ($\Delta < 0.15$), 
The latter appears for high 
 $\mu$ and $\dot{m}$, while the former occurs for low $\mu$ and $\dot{m}$.
\subsection{Width, period, and $\mu$}
\begin{figure}
\includegraphics[width=\columnwidth]{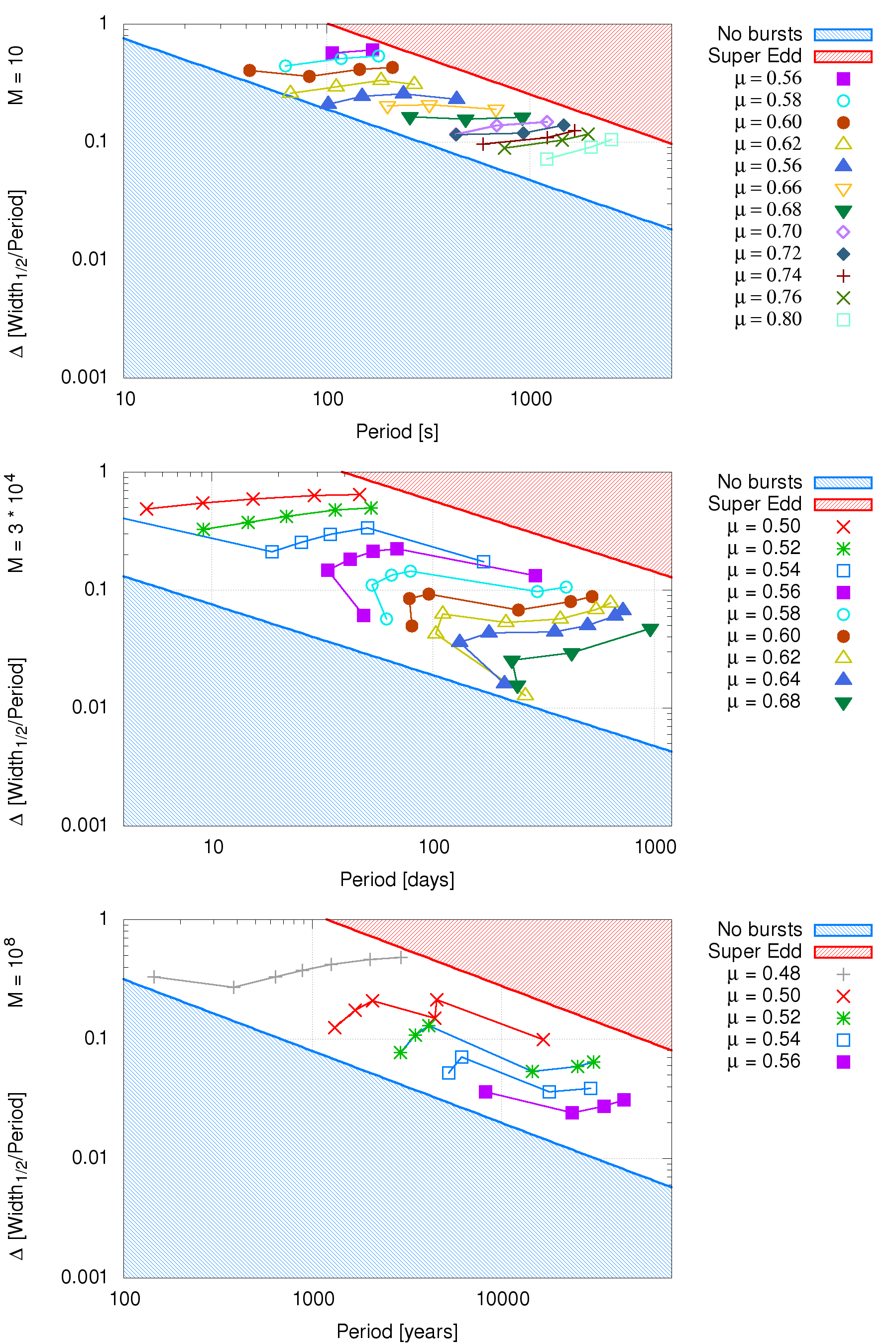}
\caption{{Upper graph}: Dependence between period and width of flares for a stellar-mass 
black hole ($M = 10 M_\odot$). The graph was prepared for a range of
$\mu \in [0.44:0.8]$.
{Middle graph}:
Dependence between period and width of flares for an intermediate-mass 
black hole ($M = 3 \times 10^4 M_\odot$). The graph was prepared for a range of
$\mu \in [0.44:0.8]$. 
{Lower graph}:
dependence between period and width of flares for a supermassive
 black hole ($M = 10^8 M_\odot$). The graph was prepared for a range of
  $\mu \in [0.44:0.56]$.
The colored lines represent isolines for different 
$\mu$.}
\label{fig:sps}
\label{fig:ips}
\label{fig:aps}
\label{fig:b}
\end{figure}

\begin{figure}
\includegraphics[width=\columnwidth]{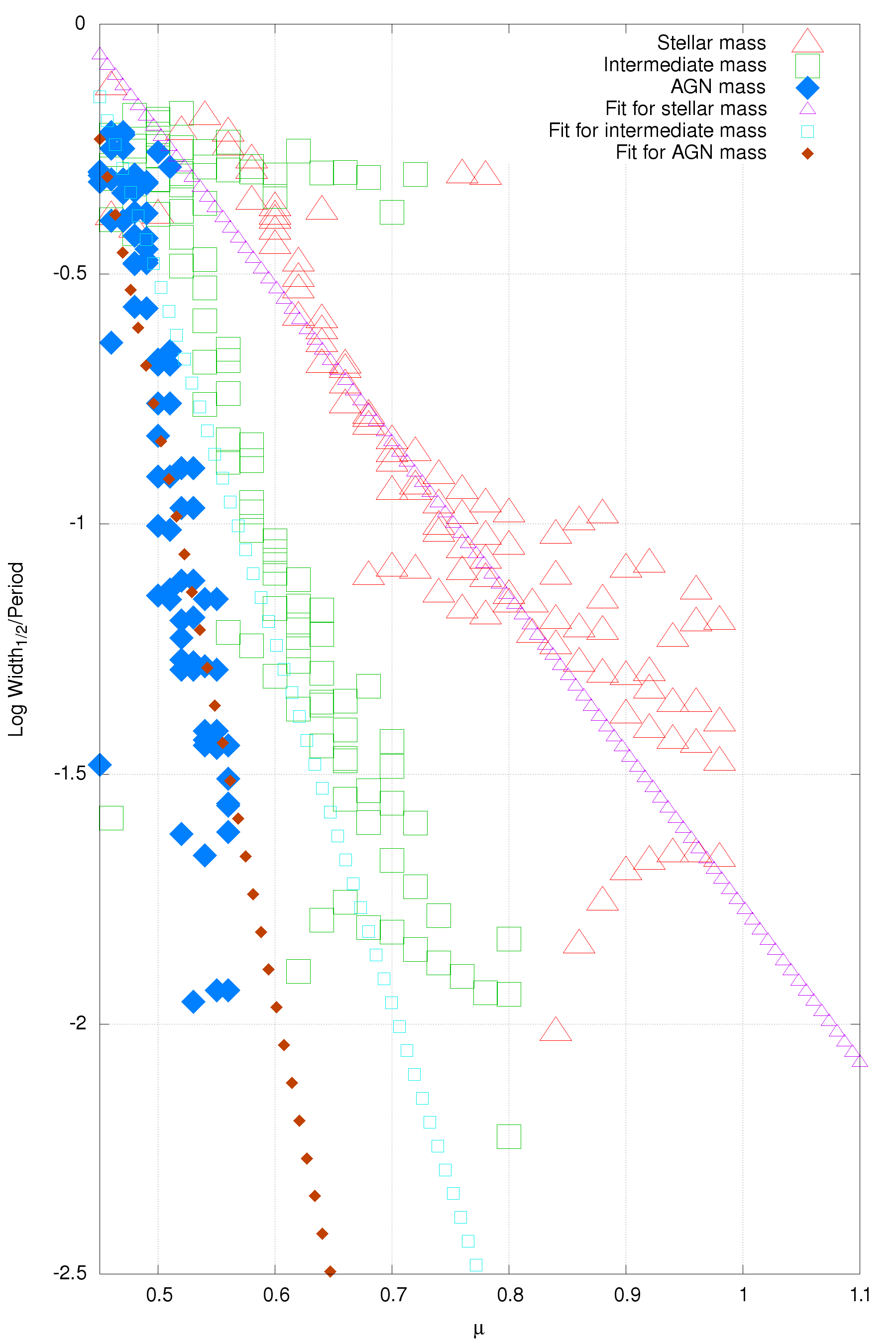}
\caption{Dependence between width, $\mu$ parameter and mass. One triangle 
represents one model. Triangles represent the model for the stellar-mass black 
hole accretion disks ($M = 10 M_\odot$), squares represent the model for
the intermediate mass black hole accretion disks ($M = 3 \times 10^4 M_\odot$)
and diamonds represent the
model for the supermassive black hole
accretion disks ($M = 3 \times 10^8 M_\odot$). Lines represent fits for each mass.}
\label{fig:s1}
\end{figure}

In Fig. \ref{fig:sps} the relation between the 
period of flares and width of flares   is presented.
The timescale of flare scales is approximately inversely proportional to the mass.
According to Fig.  \ref{fig:sps},
the flare duration to period rate $\Delta$  only
weakly depends on the value of accretion rate. The dependence on $\mu$ is more significant for all the 
probed masses. Thus, the flare shape is determined mostly by the microphysics of the turbulent 
flow and its magnetisation, which is hidden in the $\mu$ parameter, and 
not by the amount
of inflowing matter expressed by the value 
of accretion rate $\dot{m}$. 
 In Figure \ref{fig:s1}, we present the result from Fig. \ref{fig:sps}
 with linear log fits connecting the values of
$\Delta$, $\mu$, and the black hole mass $M$, which can be described by the formula:
\begin{equation}
 - \log \Delta = (1.9 + 1.2 \log M)(\mu - 3/7).
 \label{eq:deltammu}
 \end{equation}

 From Eq. (\ref{eq:deltammu}), we can conclude that the flare shape 
depends predominantly on the disk magnetisation.
{ We note that Equation (\ref{eq:deltammu})
 enables us to estimate the behaviour of the sources even for the 
values of $\mu$ higher than those used in Figure 7.
}
As a result,
for larger $\mu$ we get in general narrower flares. This effect is 
even more pronounced for larger black hole masses. Therefore, the outburst
flares are more likely to occur for larger values of $\mu$.
\subsection{Amplitude and accretion rate}
\label{amdots}
\begin{figure}
\includegraphics[width=\columnwidth]{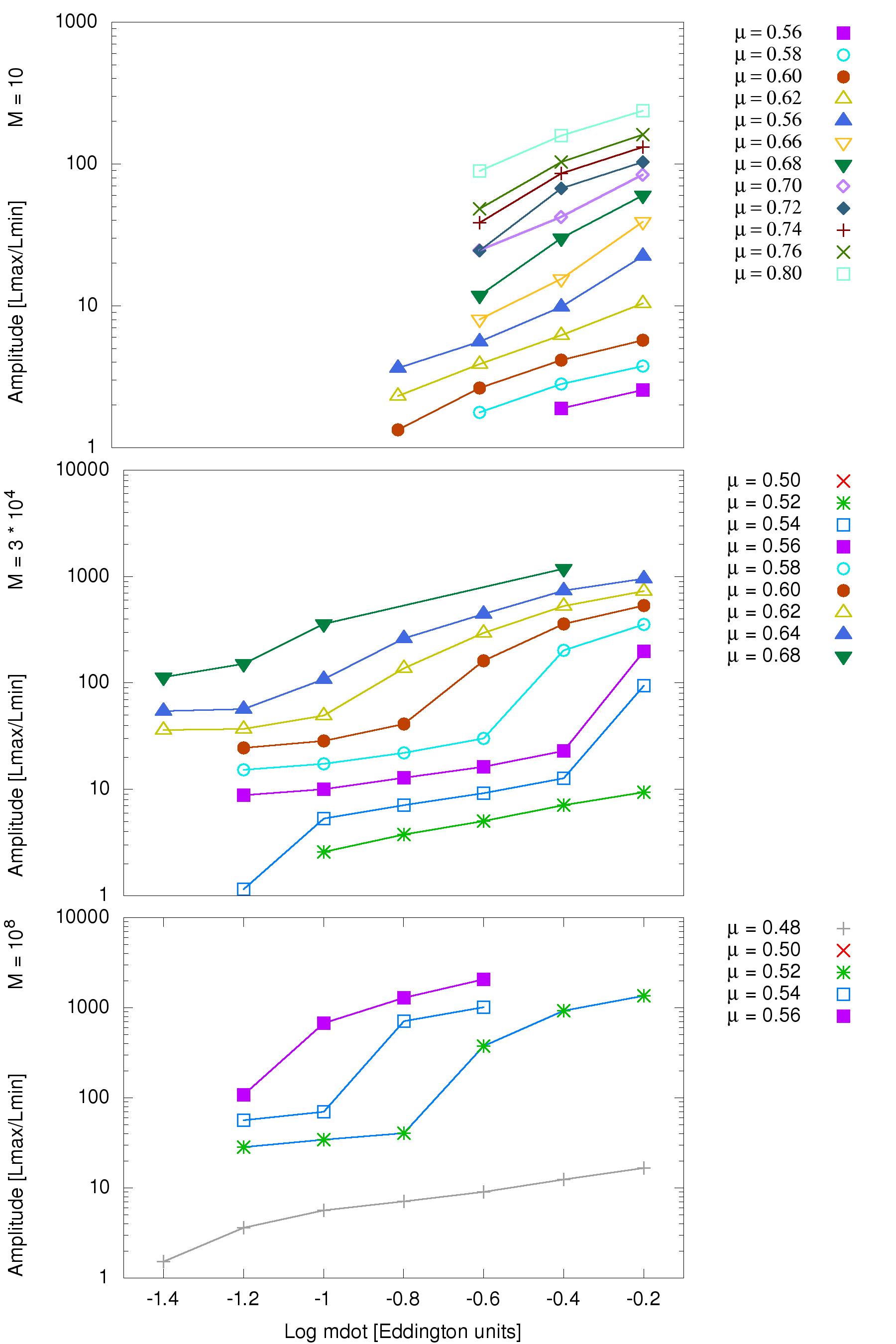}
\caption{{Upper graph:} Dependence between amplitude and accretion rate for a stellar-mass 
black hole ($M = 10 M_\odot$).{Middle graph:} Dependence between amplitude and accretion rate for an intermediate-mass 
black hole ($M = 3 \times 10^4 M_\odot$).{Lower graph:} Dependence between amplitude and accretion rate for stellar-mass 
black hole ($M = 10^8 M_\odot$). The graph was prepared for a range of
$\mu$. We can see simple unambiguous dependence for different
$\mu$-models.}
\label{fig:c}
\label{fig:samdotll}
\label{fig:amdotll}
\label{fig:aamdotll}
\end{figure}

In Figure \ref{fig:samdotll} we show the dependence between relative
amplitudes and accretion rates $\dot{m}$ for the stellar mass, intermediate mass, and supermassive black holes, respectively. We see a monotonic dependence for any value of $\mu$ and mass. Because of the nonlinearity of evolution equations
with respect to $\dot{m}$, we cannot present any simple scaling relation between $\dot{m}$ and amplitudes nor periods.
\subsection{Limitations for the outburst amplitudes and periods}
In  Figures \ref{fig:sas} and \ref{fig:sps} the dark areas mark our 
numerical estimations for the 
possibly forbidden zones in the case of microquasar
accretion disks. From those figures, we get the following fitting formulae:
\begin{equation}
  0.3 \times A^{-0.5} < \Delta < 2 A^{-0.5}
  \label{eq:ess1}
 \end{equation}
 \begin{equation}
  3 \times (P~[s])^{-0.6} < \Delta < 50 \times 16 P~[s]^{-0.6}.
  \label{eq:ess2}
 \end{equation}
 Eqs. (\ref{eq:ess1}) and  (\ref{eq:ess2}) result in the following estimation for $P$ and $A$
 \begin{equation}
   1.96 \times A^{0.83} < P ~[s] < 630 \times A^{0.83}.
    \label{eq:ess3}
 \end{equation}
In Figures \ref{fig:ias} and \ref{fig:ips} we have shown the estimated range of the possibly forbidden zones in the case of the 
accretion disks around the intermediate-mass black holes. 
From those figures, we can derive the following formulae:
 \begin{equation}
  0.07 \times A^{-0.5} < \Delta < 2.5 \times A^{-0.5}
  \label{eq:esi1}
 \end{equation}
 \begin{equation}
     0.3 (P~[{\rm days}])^{-0.6} < \Delta < 9 \times P[{\rm days}]^{-0.6}.
     \label{eq:esi2}
 \end{equation}
 Eqs. (\ref{eq:esi1}) and  (\ref{eq:esi2}) result in the following estimation for $P$ and $A$:
\begin{equation}
 0.0021 \times A^{0.83} < P < 7500 \times A^{0.83}.
     \label{eq:esi3}
\end{equation}
In figures \ref{fig:aas} and \ref{fig:aps} the dark shaded areas mark our estimations for the possibly forbidden zones for the case of
supermassive black hole accretion disks. 
We get the following formulae from those figures:
 \begin{equation}
  5 \times A^{-0.5} < \Delta < 70 \times A^{-0.5}
  \label{eq:esa1}
 \end{equation}
 \begin{equation}
  0.35 \times (P ~[{\rm years}])^{-0.6} < \Delta < 2.5 \times (P ~[{\rm years}])^{-0.6}.
  \label{eq:esa2}
 \end{equation}
Eqs. (\ref{eq:esa1}) and  (\ref{eq:esa2}) result in the following estimation for $P$ and $A$:
\begin{equation}
 4.67 \times A^{0.83} < P < 16.6 \times A^{0.83}.
 \label{eq:esa3}
\end{equation}
 \section{HLX-1 mass determination}
\label{sect:hlx1}
The grids of models deliver some information about the correlation between 
the observed light curve features and the model parameters. From the Eq. (\ref{eq:20160713}) we can determine 
the mass of an object directly from its light curve:
\begin{equation}
 M [M_\odot] = 0.45 P[s]^{0.87} A^{-0.72}.
\label{eq:massestimation}
\end{equation}
The $\Delta - \mu - M$ dependence from Fig \ref{fig:s1} and Eq. (\ref{eq:deltammu}), combined with Eq. (\ref{eq:massestimation}) gives us
the exact estimation on $\mu$ 
\begin{equation}
 \mu = 3/7 + \frac{- \log \Delta}{1.49 + 1.04 \log P - 0.864 \log A}.
 \label{eq:muestimation}
\end{equation}
The Ultraluminous X-ray sources (ULXs) are X-ray sources that exceed the Eddington limit for 
accretion on stellar-mass black holes (for $ M = 10 M_{\odot}$ 
the limit reaches $1.26 \times 10^{39} $erg$/$s).
It is thought that {some of} those objects consist of black holes with masses at the level of $100-10^6 M_{\odot}$. 
Those objects are not the product of collapse of single massive stars \citep{Davis2011}.
HLX-1 is the best known case of a ULX {being an IMBH candidate}, which has undergone six outbursts 
spread in time over several years
with; an average period of about $400$ days, a duration time of about 30-60 days, and a ratio between its
maximum and minimum luminosity $L_{\rm max}/L_{\rm min}$ of about several tens. The average
bolometric luminosity $( (\Sigma_i L_i \Delta t_i )/(\Sigma_i \Delta t_i) )$,
when $L_i$ is the luminosity at a given moment and $\Delta t_i$ is the gap between two observation points.
 The SWIFT XRT observation is at the level 
of  $1.05 \times 10^{42} \times (K/5) $  erg s$^{-1}$, 
where $K$ is the bolometric correction.
The exact value of the bolometric correction is strongly model-dependent. 
The fits of the thermal state with the \textit{diskbb} model \citep{Servillat2011}
imply a disk temperature
$T_{\rm in}$ in the range of $0.22 - 0.26$ keV, which, combined with the
0.2-10 keV flux and the distance to the source, 
implies a black hole mass of about $10^4$, if the model is used with 
the appropriate normalization, and 
a bolometric correction of $1.5$ for the $0.3 - 10$ keV spectral range. 
The use of the \textit{diskbb} model for 
larger black hole mass, $10^5$, implies an inner temperature, $T_{\rm in}$
of $0.08$ keV, much lower than observed, but then a much larger bolometric correction at $6.6$. However, 
the larger mass cannot be ruled out on the basis of the spectral analysis since it is well
known that the disk emission is much more complex than the \textit{diskbb} predictions, and in particular the inner
disk emission has colour temperature much higher than the local black body by a factor 2 - 3 
 (e.g. \citet{DoneDavis2008} and see also \citet{Sutton2017}). 
Thus the overall disk emission may not be significantly modified for the outer
radii, and the hottest tail can still extend up to the soft X-ray band. 
A factor of 5 qualitatively accounts for that trend 
\citep{Maccarone2003,Wu2016}.
In fact our method, which determines the mass, accretion rate, and parameters $\mu$ and $\alpha$ does
not depend on
the scaling factor connected with the bolometric luminosity and its validity is only indicative, 
in contradiction to P, A, and $\Delta$ estimations, that are the outcome of the non-linear internal dynamics 
of the accretion 
disk and give independent approximations on the disk parameters (e.g. see Eqs. 
(\ref{eq:massestimation}) and (\ref{eq:muestimation})).
The redshift of ESO 243-49, the host galaxy of { HLX-1,} is
equal to z=0.0243, and the distance is around 95 Mpc (depending on the cosmological constants, but the uncertainty is quite small). 
The inclination angle is not well constrained by the observations.
As we see from equation (44), the BH mass of HLX-1 is not sensitive to the luminosity but depends on the amplitude (A).
The observed values of the source variability period, $P \approx 400   $ days, and amplitude $A \approx 20$ and $\Delta \approx 0.13$, used in the Eqs.
(\ref{eq:massestimation}) and (\ref{eq:muestimation}), result in a black hole mass of
$M_{\rm BH, HLX-1} \approx 1.9 \times 10^5 $M$_\odot$
and $ \mu \approx 0.54$. These values are close to the model parameters, which are necessary to reproduce the
light curve.
Therefore, we conclude that our model of an accretion disk with the  modified viscosity law gives a proper explanation
of HLX-1 outbursts. 
 The model to data comparison, being the result of our analysis, is presented
 in the Figure \ref{fig:hlxobs}. 
  The observation presented in Fig. \ref{fig:hlxobs} and our numerical model were presented also in \citet{Wu2016}.
 In that article, the source HLX-1 was also compared to a broad ensemble of XRBs and AGNs. 
 In \citet{Wu2016} we already outlined the method presented in the current work, but here we present much a more detailed description of our generalised model.
In particular, we tested the large parameter space of our computations, and verified their influence on the observable characteristics of the sources. 
We also slightly better determined the mass of the black hole in this source.
 According to Eqs. (\ref{eq:massestimation}) and (\ref{eq:muestimation}),
 and Fig. \ref{fig:hlxobs}, taking into account the value of bolometric luminosity,
 we propose the following parameters for HLX-1: $M = 1.9 \times 10^5 M_\odot$, $\dot{m} = 0.09 - 0.18$,
 $\alpha = 0.02$ and $\mu = 0.54$.
\section{Disk flare estimations}
\label{sect:esti}
We derive the general estimations for the oscillation period $P$, its amplitude $A,$ and relative duration $\Delta$
thanks to computing a large grid of models. We achieve this despite the uncertainty of the viscosity parameter $\mu$. 
The results are presented in Table \ref{tab:fit}.

\begin{figure}
\includegraphics[width=0.95\columnwidth]{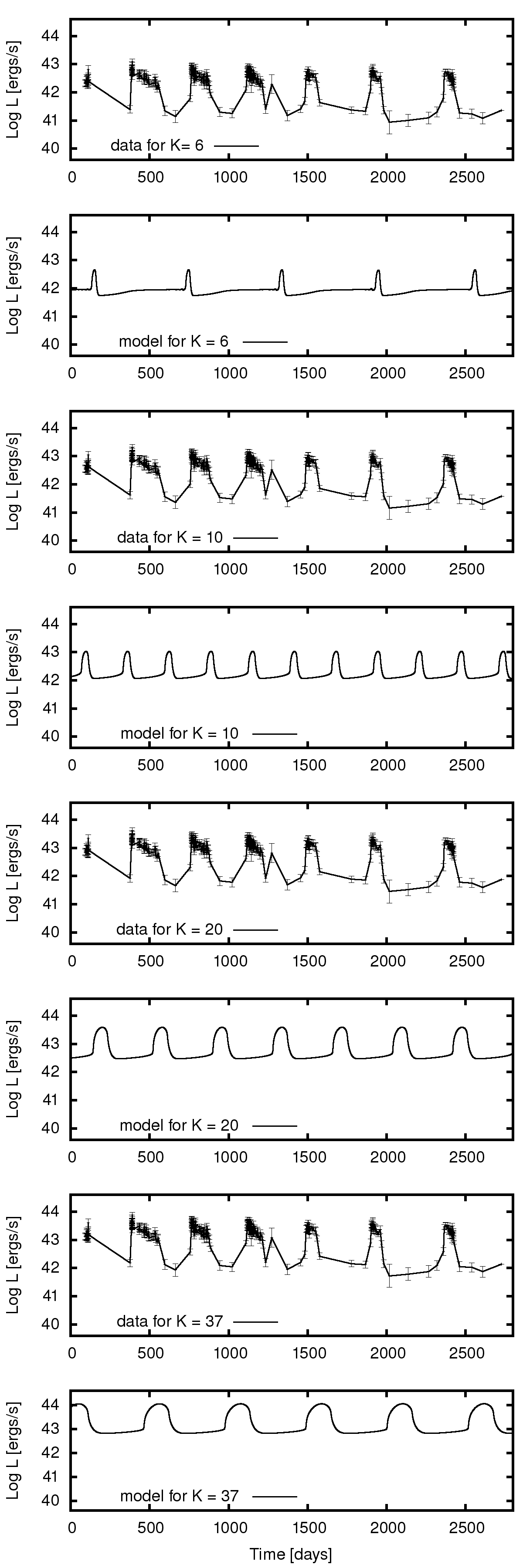}
\caption{Comparison between the SWIFT XRT observational light curve of HLX-1 and the models for different bolometric corrections K. Model parameters: $\mu = 0.54, \alpha = 0.02, \dot{m} = 0.009*K$ and $M = 1.9 \times 10^5 M_{\odot}$. }
\label{fig:hlxobs}
\end{figure}

We compare the allowed parameters of duration to period ratios
(columns $3$ and $4$), periods (columns $5$ and $6$), and flare durations (columns $7$ and $8$). We get
three values of relative amplitude for each source to present possible limitations for periods and
duration times. The amplitude $A = 2$ represents faint flares, 
like those 
in the microquasar IGR J17091 \citep{Janiuk2015}. The amplitude $A =10$
corresponds to a more developed instability case, like that in the $\rho$ states of object GRS1915 
\citep{GRS2000Belloni}. The amplitude $A = 100$ is connected with the huge bursts, as in HLX-1 presented in the
Figure \ref{fig:hlxobs}. 
The timescales presented in the Tables are on the order of minutes for the microquasars, of months for the 
HLXs, and of millenia for the AGN, which correspond to their viscous timescale. 
The period is strongly dependent on the amplitude and can change by several orders
of magnitude for each mass. The estimations given here were made for the values of $\mu$ which are sufficient for the outbursts, but do not
determine exactly the value of this parameter.   
Figure \ref{fig:13} (see next Section) presents a universal dependence between duration times and bolometric
luminosities for the X-ray sources of different scales  (microquasars with $M = 10 M_\odot$,
IMBHs with $M =  3 \times 10^4 M_\odot$, and AGN with $M = 10^8 M_\odot$).
 Exact duration values from the fit presented in Eq. $(\ref{eq:duration})$ are shown in Table \ref{tab:fita}.
 For microquasars, the values included in Table \ref{tab:fit} correspond to the typical values for small and intermediate flares, the same for the case of IMBHs. In case
 of AGN, the appearance of big flares is necessary to verify our model with the observational data  presented in the Figure \ref{fig:13}.
  \begin{table}
  \begin{tabular}{|c|c|c|c|}
  \hline
  \multicolumn{4}{|c|}{Flare characteristic quantities for microquasars:}
\\ \hline
   A &  $\Delta$ &  $P$ [s] & $P \Delta$ [s]  
  \\ \hline
  *2  & 0.212 - 1 & 3.48 - 1120 & 0.739 - 1120
  \\ \hline
  **10 & 0.0949 - 0.632 & 13.3 - 4260 & 1.26 - 2690
  \\ \hline
  ***100 & 0.03 - 0.2 & 89.6 - 28800 & 2.69 - 5760
  \\ \hline
   \multicolumn{4}{|c|}{Flare characteristic quantities for IMBHs:}
  \\ \hline
   A &  $\Delta$   & $P$ [days]&  $P \Delta$ [days]  
  \\ \hline
  *2 & 0.0494 - 1 & 0.533 - 5330 & 0.00264 - 5330
  \\ \hline
  **10 & 0.0221 - 0.79 & 0.203 - 20282 & 0.00449 - 16030
  \\ \hline
  ***100 & 0.007 - 0.25 & 1.37 - 13700 & 0.0960 - 34300
  \\ \hline
     \multicolumn{4}{|c|}{Flare characteristic quantities for AGNs:}
  \\ \hline
   A & $\Delta$  &  $P$ [years] &  $P \Delta$ [years]
  \\ \hline
  *2 & 0.248 - 1 & 24.9 - 12100 & 6.16 - 21100 
  \\ \hline
  **10 & 0.111 - 0.796 & 94.7 - 46500 & 10.5  - 36300
  \\ \hline
  ***100 & 0.0035- 0.25 & 640 - 311000 & 22.4 - 77700
  \\ \hline
  \end{tabular}
\caption{Outburst duration values for three kinds of source. The values for microquasars are
expressed in seconds, values for IMBHs in days and values for AGNs in years. Flare regimes * - small flicker, 
** - intermediate, *** - burst}
\label{tab:fit}
 \end{table}
  \begin{table}
  \begin{tabular}{|c|c|c|c|}
 \hline
  Source & Mass [$M_{\odot}$] & *$P \Delta_{0.1}$  & **$P \Delta_1$
  \\ \hline
  Microquasar  & $10$ &$33 $s &$ 595$ s
  \\ \hline
  IMBH  & $3 \times 10^{4}$ days& $8.59$ &$153$ days
  \\ \hline 
  AGN & $10^{8}$ & $596$years &$10603$ years
  \\ \hline
  \end{tabular}
\caption{Outburst duration values for three kinds of source. The values are taken from Eq. $(\ref{eq:duration})$.
* - duration for $L = 0.1 L_{\rm Edd}$, ** - duration for $L =  L_{\rm Edd}$}
\label{tab:fita}
 \end{table}
 \section{Summary and discussion}
\label{sect:discussion}

In this work, we studied the accretion disk instability induced by the dominant 
radiation pressure, with the use of the generalised prescription for the
stress tensor. We adopted a power-law dependence, with an index $\mu$, to describe the contribution 
of the radiation pressure to the heat production. 
In other words, the strength of the radiation pressure instability 
deepens with growing $\mu$.
We computed a large grid of time-dependent 
models of accretion disks, parameterised by the black hole mass, and mass accretion rate.
We used the values of these parameters, which are characteristic
 for the microquasars, intermediate black holes, or AGN. 
One of our key findings is that this model can be directly applicable 
for determination of the black hole mass and accretion rate values, for example, 
for the Ultraluminous X-ray source HLX-1, and possibly also for other sources. 
We also found that the critical accretion rate, for which the thermal instability appears, decreases with growing $\mu$
 (see Figure \ref{fig:sa}).
Also, the amplitudes of the flares of accretion disks in AGN are larger than the amplitudes of flares
in microquasars and in IMBHs. The flare period grows monotonously with its amplitude, for any value of mass
(see Figure \ref{fig:sap}). The outburst width remains in a well-defined relationship
with its amplitude (see Figure \ref{fig:sas}). 
We also found that there is a significant negative correlation between $\mu$ and the ratio of the flare duration to the variability period, $\Delta$. 
On the other hand, the dependence between the outburst amplitude $A$ and the mass 
accretion rate $\dot{m}$ is non-linear and complicated. 
Our results present different variability modes (Figures \ref{fig:ifl} and \ref{fig:iou}).  The flickering mode is presented in Fig. \ref{fig:ifl}. In this mode the relative amplitude is small, and flares repeat after one another.
In the burst mode the amplitude is large, and the maximum luminosity can be hundreds of times greater
than minimal. An exemplary light curve is shown in
Fig. \ref{fig:iou}. In this mode we observe long separation between the flares (i.e. an extended low luminosity state), dominated
by the diffusive phenomena. A slow rise of the luminosity is the characteristic property of the disk instability model.
 \begin{table}
  \begin{tabular}{|c|c|c|c|c|c|c|}
  \hline
  {\bf Source} &  {\bf ID} & {\bf P} & {\bf A} & {\bf $\Delta$} & {\bf $\frac{M}{M_\odot} (\frac{\alpha}{0.02})^{1.88}$} & {\bf $\mu$}*
   \\ \hline
   IGR &    $\nu_{I} $  & $45$s & $2.5$ & $0.15$ & $6.38$  & $0.717$
   \\ \hline
   IGR &   $\rho_{I A}$  & $30$s & $3.5$ & $0.3$ & $3.52$ & $0.634$
      \\ \hline
   IGR &    $\rho_{I B}$  & $30$s & $4$ & $0.4$ & $3.198$ & $0.589$
      \\ \hline
   GRS &   $\nu_{G} $  & $90$s & $4$ & $0.1$ & $8.31$ & $0.763$
      \\ \hline
   GRS &  $\rho_{G A} $  & $45$s & $5$ & $0.25$ & $3.87$ & $0.661$
      \\ \hline
   GRS &  $\rho_{G B} $  & $40$s & $4.5$ & $0.4$ & $3.77$ & $0.583$
      \\ \hline
   HLX &   -  & $400$d & $2.5$ & $0.14$ & $1.88 \times 10^5$ & $0.534$
      \\ \hline\hline
   AGN${*}$ &   -  & $10^5$y & $100$ & $0.1$ & $1.6 \times 10^8$ & $0.515$   
  \\ \hline
  \end{tabular}
  \caption{Characteristic quantities of the RXTE PCA light curves for Galactic sources presented in
\citep{Altamirano11a} (columns $4,5,6$) supplemented with HLX and AGNs, and estimations of the {$mass-\alpha$ relations}
and magnetisation of 
sources (columns $7,8$). 
{\bf Notation:} IGR = IGR J17091, GRS = GRS1915, HLX = HLX-1, AGN - typical value for the sample of AGNs presented 
in \citep{Czerny2009}. The OBSIDs of the ligth curves are as follows:
 $\nu_{I} =96420-01-05-00$ ( $\nu$ state), $\rho_{I A} =96420-01-06-00$ ({ $\rho$ state}), $ \rho_{I B} =  96420-01-07-00$({ $\rho$ state})
 ,$\nu_{G}  =10408-01-40-00$ ( $\nu$ state) ,  $\rho_{G A} = 20402-01-34-00$ ({ $\nu$ state}) and  $\rho_{G B} = 93791-01-02-00$ ({ $\rho$ state}), * - Values for a typical AGN.}

\label{tab:deter}
\end{table} 

\subsection{Mass - $\alpha$ relation}
\label{sect:ma}

\begin{figure}
\includegraphics[width=\columnwidth]{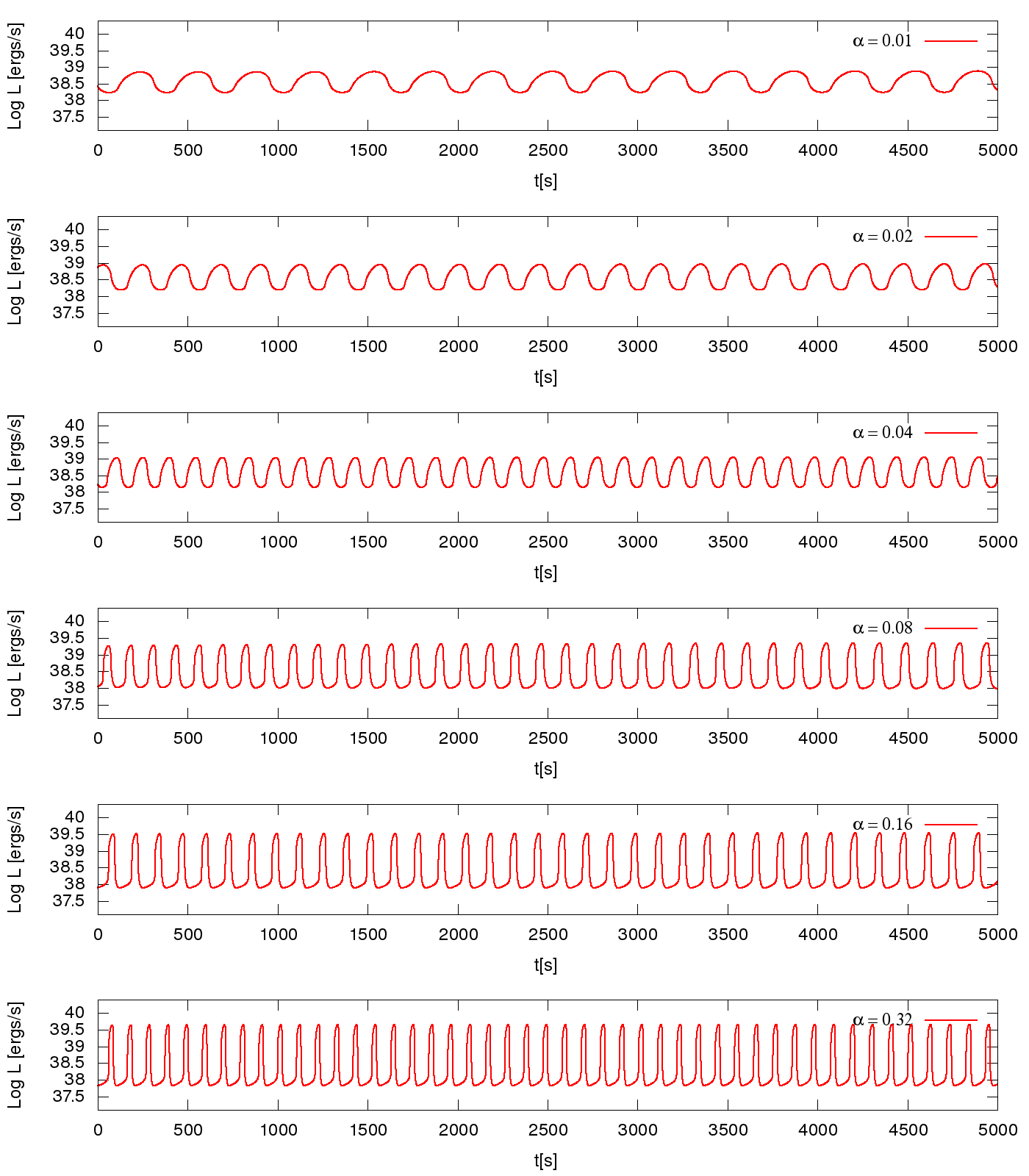}
\caption{Light curves for six different values of $\alpha$ for $M = 10 M_{\odot}$, $\dot{m} = 0.64$, and $\mu = 0.6$. }
\label{fig:lca}
\end{figure}

\begin{figure}
\includegraphics[width=\columnwidth]{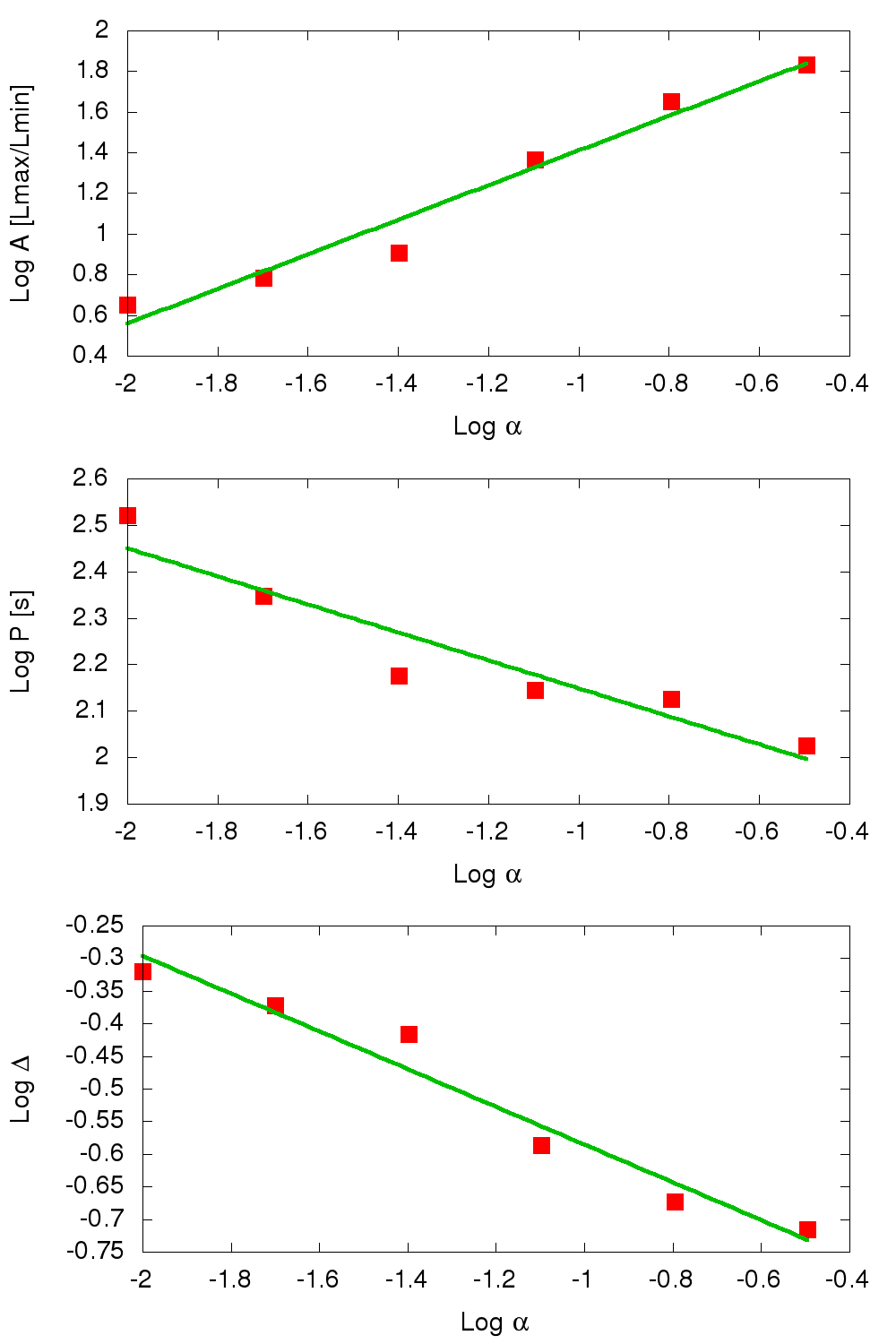}
\caption{Dependence between the $\alpha$ parameter and observables for $M = 10 M_{\odot}$, $\dot{m} = 0.64$, and $\mu = 0.6$ for $\alpha \in [0.01:0.32]$.}
\label{fig:oa}
\end{figure}

Since the thermal and viscous timescales strongly depend on $\alpha$, which has only
ad-hoc character \citep{King2012} and  does not constitute any fundamental physical 
quantity, $\alpha$ is the parameter describing development of the MHD turbulence in the accretion 
disk. Thus $\alpha$ should, to some extent, vary depending on the source source and its state; for example, the value of $\alpha$ for the AGN accretion disks can differ from its value for the disks in
X-ray binaries. In Fig.  $\ref{fig:lca}$ we present different light curve shapes for six 
different values of $\alpha$. In Fig. $\ref{fig:oa}$ we present the dependence of the light curve
observables on $\alpha$.
{
The formulae describing fits in Fig. $\ref{fig:oa}$ are as follows:}
\begin{equation}
 \log O_X = b_X \log \alpha + c_X
\label{eq:ox}
,\end{equation}
{
where $O_X = A$, P$ $[in seconds], or $\Delta$. The coefficients are as follows $b_A = 2.25 \pm 0.11$,
$c_A = 0.85 \pm 0.05$, $b_P = -0.3 \pm 0.05$, $c_P = 1.85 \pm 0.07$, $b_{\Delta} = -0.29
\pm 0.03$, $c_{\Delta} = -0.87 \pm 0.05$. 
According to Eqs. $(\ref{eq:massestimation})$ and $(\ref{eq:muestimation})$ we obtain from Eq. $(\ref{eq:ox})$ the following:
}
\begin{equation}
 M [M_\odot] = 0.45 P[s]^{0.87} A^{-0.72} (\frac{\alpha}{0.02})^{1.88},
\label{eq:mea}
\end{equation}
\begin{equation}
 \mu = 3/7 + \frac{- \log \Delta + 0.87 \log (\frac{\alpha}{0.02})}{1.49 + 1.04 \log P - 0.864 \log A}.
 \label{eq:muea}
\end{equation}
\subsection{Radiation pressure instability in microquasars}
Quantitatively, our numerical computations, as well as the fitting 
formula (\ref{eq:20160713}), give the adequate description
of the characteristic  `heartbeat' oscillations of the two known 
microquasars: GRS 1915+105, and 
IGR 17091-324. Their profiles resemble those observed in the so-called $\rho$ state of these sources, as found, for example, on 27th May, 1997 \citep{Pahari14}. 

For the microquasar IGR J17091, 
the period of the observed variability is less than $50$s, as observed in the most regular
heartbeat cases, that is, in the $\rho$ and $\nu$ states \citep{AltamiranoBelloni2011}. 
The $\nu$ class is the second most regular variability class after $\rho$, much more 
regular than any of the other ten classes described in \citet{GRS2000Belloni} ($\alpha$, $\beta$,$\gamma$, $\delta$, $\theta$, $\kappa$,
$\lambda$,$\mu$, $\phi$, $\chi$).

The $\rho$ state is sometimes described as `extremely regular' \citep{GRS2000Belloni},
with a period of about $60-120$ seconds for the case of GRS1915. The class $\nu$ includes
typical Quasi-Periodic Oscillations with relative amplitude large than $2$ and a period of
$10-100$s. 

 We apply the results of the current work to model the heartbeat states qualitatively. 

Eqs. $(\ref{eq:20160713})$ and $(\ref{eq:deltammu})$ allow us to determine the values of
BH masses for the accretion disks and the $\mu$ parameter. The results are given in Table \ref{tab:deter}. For the 
$\rho$-type light curves we can estimate the $mass-\alpha$ $(\frac{M}{M_\odot} (\frac{\alpha}{0.02})^{-1.88})$ parameter of IGR J17091-3624
at the level of $3.2-3.5$ and GRS 1915+105 at the level of $3.7-3.9$.  $\mu = 0.58 - 0.63$
for the IGR J17091-3624 and $0.58 - 0.66$ for the $GRS1915,$ respectively. From the $\nu$-type 
light curves we get significantly larger values of $M-\alpha$ parameters and $\mu$s.

Our model thus works properly for the periodic and regular oscillations,
which are produced in the accretion disk for a broad range 
of parameters, if only the instability appears.
 Irregular variability states $\alpha$, $\beta$, $\lambda$ and
$\mu$ should be regarded as results of other physical processes. 
The explanation of class $\kappa$ of the microquasar GRS 1915 
variable state, \citep{GRS2000Belloni} which presents modulated QPOs, seems to be on the border of applicability.
In general, the method is correct for estimation of the order of magnitude, although
not perfect for exact determination of the parameters due to the nonlinearity of the model.
For this paper we assumed a constant value of $\alpha = 0.02$ which could not be true for all
values of masses.
For a source with known mass, such as GRS1915 \citep{Greiner2001,Steeghs2013}, we can use previous estimations as a
limitation for the value of $\alpha$,  as presented in Section \ref{tab:6}. \citet{Steeghs2013} estimated the mass of GRS1915 at the level
$10.1 \pm 0.6 M_\odot$. From the high-frequency QPO comparison method used by \citet{Rebusco2012} we know 
the GRS/IGR mass ratio, which is at the level of 2.4.
Combining the
results of \citet{Rebusco2012} and later the GRS1915 mass estimation from \citet{Steeghs2013}, for the IGR J17091 
we get $M = 4.2 \pm 0.25 M_{\odot}$. Results of \citet{IyerNandi2015} suggest the probable mass range of IGR J17091
is between $8.7$ and $15.6 M_{\odot}$.
 \begin{table}
  \begin{tabular}{|c|c|c|c|c|}
  \hline
   {\bf source} &  {\bf ID} &   {\bf *$\mathcal{M}$} & $\frac{M}{M_\odot}$ & $\alpha$ 
   \\ \hline
   IGR &    $\nu_{I} $ &  $6.38$  & $3.95 - 4.45$** & $0.0155 - 0.0165$
   \\ \hline 
   IGR &   $\rho_{I A}$  &  $3.52$ & $3.95 - 4.45$** & $0.0213 - 0.0227$
      \\ \hline 
   IGR &    $\rho_{I B}$ &  $3.198$ & $3.95 - 4.45$** & $0.0223 - 0.0238$
      \\ \hline
   IGR &    $\nu_{I} $ &  $6.38$  & $8.7 -  15.6$*** & $0.0235 - 0.0321$
   \\ \hline 
   IGR &   $\rho_{I A}$  &  $3.52$ & $8.7 - 15.6$*** & $0.0323 - 0.0441$
      \\ \hline 
   IGR &    $\rho_{I B}$ &  $3.198$ & $8.7 - 15.6$*** & $0.0330 - 0.0446$
      \\ \hline
   GRS &   $\nu_{G} $ &   $8.31$ & $9.5 - 10.7$ & $0.0214 - 0.0228$
      \\ \hline 
   GRS &  $\rho_{G A} $ & $3.87$ & $9.5 - 10.7$ & $0.0322 - 0.0343$
      \\ \hline
   GRS &  $\rho_{G B} $ &  $3.77$ & $9.5 - 10.7$ & $0.0327 -0.0348$
      \\ \hline
    \end{tabular}
    \label{tab:6}
    \caption{{Determination of $\alpha$ values based on the known IGR and GRS mass values \citep{Rebusco2012,Steeghs2013,IyerNandi2015} and
    mass-$\alpha$ relation presented in Table $\ref{tab:deter}$. Descriptions of the sources,
    their states and OBSIDs are presented in Table $\ref{tab:deter}.$
    * - Mass - $\alpha$ factor $(\frac{M}{M_\odot} (\frac{\alpha}{0.02})^{-1.88})$. ** - IGR mass estimation from \citet{Rebusco2012} and \citep{Steeghs2013} *** - IGR mass
    estimation from \citet{IyerNandi2015}.}}
  \end{table}
{From Tables \ref{tab:deter} and \ref{tab:6} we conclude the possible dependence between $\alpha$ and the variability 
state or the source type. For the $\nu$ state of IGR we found $\alpha \approx 0.0155 - 0.0165$, for the $\rho$ state of this source 
$\alpha \approx 0.021 -0.024$, for the $\nu$ state of GRS $\alpha \approx 0.021 - 0.023$ and for the $\rho$ state of GRS
$\alpha \approx 0.032 -0.035$.}
 {Those values can change if we assume the BHs spin, which can be near to extreme 
 in the case of GRS1915 \citep{Done2004} and very low, even retrograde in the case of IGR J17091 \citep{Rao2012}.}
  In our current model we neglect the presence of the accretion disk corona.
  {If we follow the mass estimation done by \citet{IyerNandi2015}, we get quite a consistent model for both microquasars'
  $\nu$ and $\rho$ variability states - $\alpha \approx 0.023$ for the $\nu$ state and $\alpha \approx 0.033$ the $\rho$ state,
  if only we assume the mass of IGR at the level of $9-10 M_{\odot}$, that is, 
  close to the lower limit from results of \citet{IyerNandi2015}.}
  
  {We get the above results under the assumption of negligible influence of coronal emission on the
  light curve.}  According to \citet{2006MNRAS.372..728M}, power fraction $f$
emitted by the corona is given by following formula: 
\begin{equation}
 f = \sqrt{\alpha} \Bigg[ \frac{P}{P_{\rm gas}} \Bigg]^{1 - \mu}.
\end{equation}
For our model with $\alpha = 0.02$, the values of $f$ are low, where $f = 0.141 (\frac{P}{P_{\rm gas}})^{1-\mu}$, which for 
the values of $\mu$ investigated in the paper fulfills the inequality of 
$0.0125 < f <  0.141$, if we assume a threshold maximal 
value of the gas-to-total pressure rate $\beta = \frac{P_{\rm gas}}{P}$ from Eq.(~\ref{eq:muszusz}). According to the fact 
that the (\textit{heartbeat}) states are strongly radiation-pressure dominated and the coronal emission rate is lower for lower values of
the $\mu$ parameter, (which are more likely to reproduce the observational data), we can regard the coronal emission as negligible.
\subsection{Disk instability in supermassive black holes}
\label{wykresy:a}
In the scenario of radiation pressure instability, with a considerable supply of accreting matter, 
the outbursts should repeat regularly every $10^4-10^6$ years
\citep{Czerny2009}. 
From the grid of models performed in  \citep{Czerny2009}, done for $\mu = 0.5$ 
($\tau_{ r \phi} = \alpha \sqrt{P P_{\rm gas}}$) and $10^7 M_\odot < M < 3 \times 10^9$,
they obtained the following formula expressing correlation between the duration time, $\alpha$ 
 parameter and bolometric luminosity $L_{\rm bol}$:
 \begin{equation}
\log (\frac{T_{\rm dur}}{\rm yr}) \approx  1.25 \log (\frac{L}{{\rm erg s}^{-1}}) + 0.38 \log (\frac{\alpha}{0.02}) + 1.25 \log K - 53.6
\label{eq:duration}
\end{equation}
which, for the special case $\alpha = 0.02$ and neglecting the bolometric correction, has the following form: 
\begin{equation}
\log \frac{L_{\rm bol}}{ {\rm erg s}^{-1}} = 0.8 \log (T_{\rm dur}/s) + 42.88.
\label{eq:durationa}
\end{equation}
The formula $(\ref{eq:durationa})$ also found
its confirmation in observational data for  different scales of BH masses, as presented in Fig. \ref{fig:13}. 
 This applies despite the assumption of $\mu = 0.5$ since the expected dependence on 
$\mu$ is weak. If we combine Eqs. (\ref{eq:20160713}) and (\ref{eq:deltammu}), 
we get: 
\begin{equation}
 \log (T_{\rm dur}/s) = (1.15 - 1.2(\mu - 3/7)) \log M + 0.83 \log A - 1.9\mu + 0.83.
 \label{eq:160824}
\end{equation}
From Fig. \ref{fig:13} we can suggest the approximate
dependence 
\begin{equation}
 \log A = 0.4 \log M + 0.25
  \label{eq:160824a}
,\end{equation}
since for the same model input parameters
(e.g. $\log \dot{m} = -0.2$, $\alpha = 0.02$ and $\mu = 0.56$) 
the amplitude could be even a hundred times larger for the case of AGNs
than for microquasars. 
Combining Eqs. (\ref{eq:160824}) and (\ref{eq:160824a}) and adopting $L_{bol} = \dot{m} L_{Edd, \odot}$ where
$L_{Edd, \odot} = 1.26 \times 10^{38}$erg s$^{-1}$, we get:
\begin{equation}
 \log (T_{\rm dur}/yr) = (1.91 - 1.2 \mu )(\log L- \log L_{edd \odot} - \log \dot{m} ) - 6.68.
   \label{eq:160824b}
\end{equation}
{ The Eq. (\ref{eq:160824b}) can be inverted:
\begin{equation}
 \log L = \frac{1}{1.91 - 1.2 \mu} \log (T_{\rm dur}/yr)+ 37.1 + \frac{6.68}{1.91 - 1.2 \mu} + \log \frac{\dot{m}}{0.1}.
 \label{eq:56}
\end{equation}
The above Equation is a generalised version of the results from \citet{Czerny2009} and \citet{Wu2016}.
\begin{figure}
\includegraphics[width=\columnwidth]{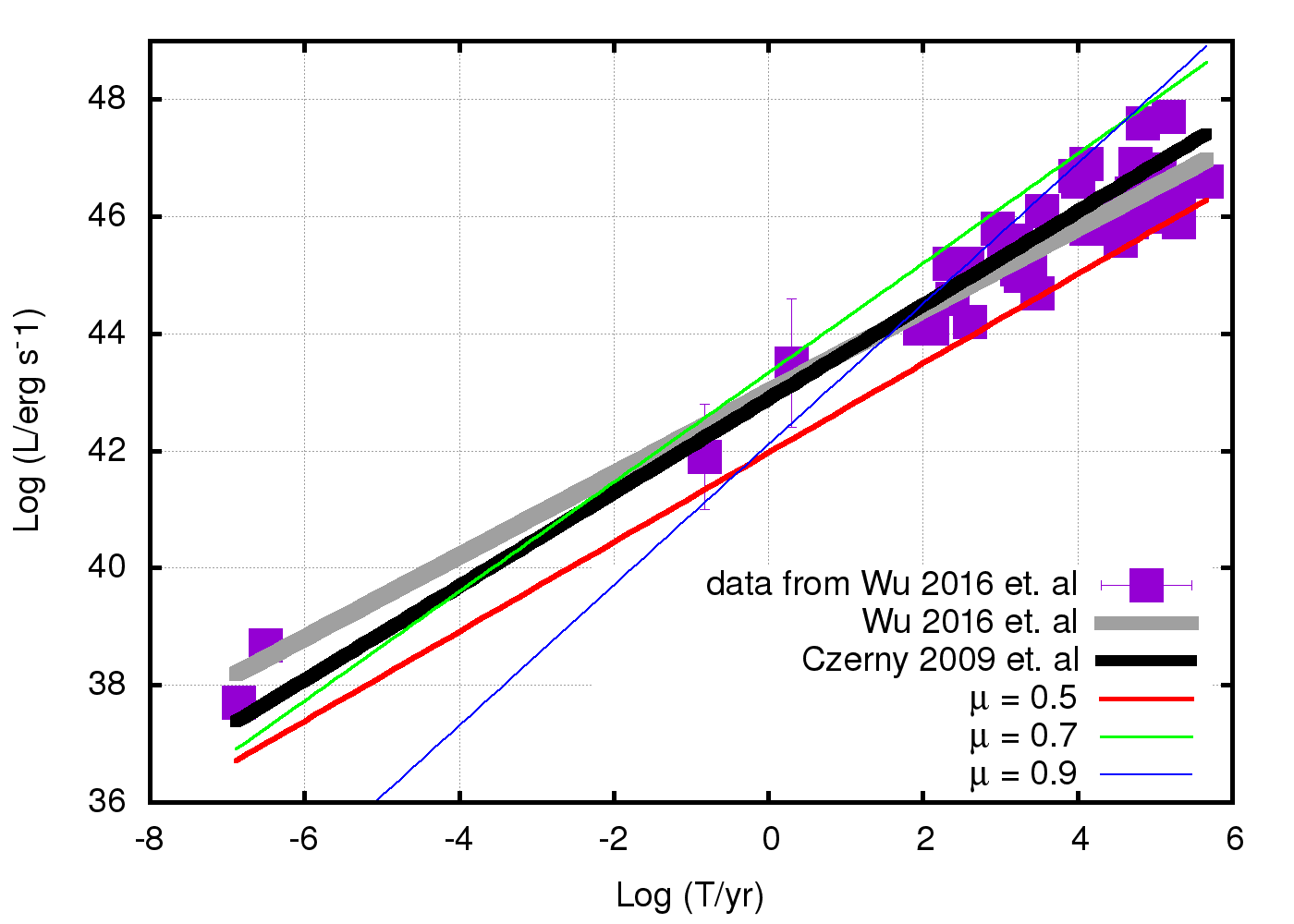}
\caption{Correlation between the bolometric luminosity and the outburst duration for different-scale BHs. Thick lines represent
the best fit from \citet{Wu2016}, and the prediction from \citet{Czerny2009}. Thin lines represent Eq. (\ref{eq:56}) for several different values of $\mu$, assuming $\dot{m} = 0.1$.}
\label{fig:13}
\end{figure}
In Fig. \ref{fig:13} we present the observational points from \citet{Wu2016}, and the theoretical lines as a result from Eq. (\ref{eq:56}) for several values of $\mu$. 
 We assumed that the Eddington accretion rate and the accretion efficiency are roughly independent for different BH masses.
The proportionality coefficient in Eq.~\ref{eq:duration} 
changes from 1.25 to $1.91 - 1.2 \mu$, that is, 1.19 for $\mu = 0.6$. 
For most of the known AGNs, except for the Low Luminosity AGNs, their luminosity in Eddington units 
is over $0.02$ \citep{AGN2006Hardy}, and the sources remain in their
 soft state, so the radiation pressure instability model should apply. 
The weak sources claimed to be AGNs, such as NGC4395 and NGC4258 \citep{LasotaNGC42,FilippenkoNGC43}
are claimed to be in the hard state, being not described accurately by the accretion disk model of Shakura-Sunyaev.
  It should be possible to study the evolution of those sources statistically.
    Based on the known  masses, accretion rates, and timescales for AGN, 
  the luminosity distribution for the samples of AGN with similar masses or
  accretion rates can be acquired. This should allow us to reproduce
 an average light curve for a range of masses and accretion rates for a survey of the known AGNs \citep{Wu2009}. The averaged light curve for a big ensemble of AGN will 
 help us to provide expected luminosity distributions or luminosity-mass, luminosity-duration relations for the AGNs existing in the universe. However, high-amplitude outbursts may complicate the study since the detection of the sources between 
 the flares may be strongly biased as the sources become very dim.  Existence of the likely 
 value of $\mu \approx 0.6$, proven  by comparison of Eqs.  (\ref{eq:duration})  and (\ref{eq:160824b}), could 
 also help in mass determination of newly discovered objects.
 Another interesting situation comprises the so-called Changing-Look-AGN, such as IC751 \citep{ricci2016}.
 Although most AGN have a variability timescale on the order of thousands of years, the shape of model light curves (sharp and rapid 
 luminosity increases) could suggest that, for some cases, luminosity changes can be observed. 
\subsection{Explanation of the HLX-1 light curve irregularity}
\label{sect:irreg}
The light curve of HLX-1, presented in the upper panel of Fig. \ref{fig:hlxobs}, despite very regular values of peak luminosity
($\log \frac{L}{{\rm erg   s}^{-1}}$ between $42.5$ and $42.6$), presents significant variability of the flare duration. According to our model, for any constant
input parameter (mass, Eddington rate, $\alpha$ and $\mu$), period, duration, and amplitude should remain constant. To our knowledge, the only explanation for such a phenomenon is variation in the input parameters. The variability of the central object mass is
too faint (approx. $10^{-8}$ per one cycle for any accreting source) to be visible. The variability of $\dot{m}$ is possible, bearing in mind the fact that accretion rate of HLX-1 (order of $10^{-3} M_\odot$ per one duty cycle) can be significantly disturbed by the tidal disruption of the minor bodies such as planets with mass ranging from $10^{-6} M_\odot$ to $10^{-3} M_\odot$.
A detailed description of this process can be found in \citet{EvansKochanek1989} and
\citet{DelSanto14} presented its application for the case of phenomena inside the globular clusters. The Eddingtion rate $\dot{m}$ is a global parameter, strongly connected with the accretion disk neighbourhood ($\dot{m}$ can change rapidly
in the case of tidal disruption). In contradiction, $\mu$ and $\alpha$ are the local parameters approaching the MHD turbulence. As $\alpha$ can be connected with the rate of the typical velocity of turbulent movement to
the sound speed \citep{1973A&A....24..337S}, $\mu$ can represent the magnetisation of the disk, as shown by the 
Eq. (\ref{eq:mumagnetic}).
 In the HLX-1 observation, out of the four observables, only the $\Delta$ parameter was changing significantly between different
flares. According to Eq. $(\ref{eq:muea})$ this follows from changing $\mu$. Specifically, the growth of $\mu$ from $0.48$ to $0.56$  is responsible
for  $\Delta$ decreasing from $0.4$ to $0.1$ . According to those results, $\mu$ was growing during the sampling time, which can be explained by a decrease in disk magnetisation.
\subsection{Conclusions}
\label{sect:discconc}
 We propose a possible application of the modified viscosity model as a description of  a regular variability pattern (heartbeat states) of black hole accretion disks 
 for the microquasars, IMBHs and AGNs. The model works for optically thick, geometrically thin disks and determines the range and scale of the radiation pressure instability. The parameter
 $\mu$, which describes viscosity, can reproduce a possibly stabilising influence of the strong magnetic field in the accretion disk. 
 Nonlinearity of the models causes appearance of different modes of the disk state (stable disk, flickering,
 outbursts).  Thanks to the computation of  computing a large grid of models we are able to present 
quantitative estimations for the variability periods 
 and amplitudes, and our model light curves reproduce several different variability patterns.  Also, many observables, such as, $L$, $P$, $A$, and $\Delta$, 
 can be used  directly to determine the physical parameters, like $\alpha$, $\mu$, $M$,
 and $\dot{m}$. Finally, our model can be successfully applied to the mass and accretion rate determination for the intermediate black 
 hole HLX-1 at the 
 level of $1.9 \times 10^5 $M$_\odot$ and $0.09 - 0.18$ respectively, updating the result from \citet{Wu2016}. The prospects of further applications to microquasars and AGNs 
are promising.
\section*{Acknowledgments}
We thank Petra Sukova for helpful discussions, Conor Wildy for the language corrections and the anonymous referee for constructive comments. 
This work was supported in part by the grants DEC-2012/05/E/ST9/03914
and 2015/18/M/ST9/00541 from the Polish National Science Center.
\bibliographystyle{aa}
\bibliography{viscb} 
\end{document}